\documentclass[twocolumn]{aastex63}

\usepackage{comment}
\usepackage{lipsum}
\usepackage{threeparttable}
\usepackage{xcolor}
\usepackage{colortbl}
\usepackage{color}
\usepackage{array}
\usepackage{multirow}
\usepackage{bigstrut}
\usepackage{amsfonts}
\usepackage{amsmath}
\usepackage{amssymb}
\usepackage[pangram]{blindtext}
\usepackage{tabu}
\usepackage{threeparttable}
\usepackage{ mathrsfs }
\usepackage{textcomp}
\colorlet{mycolor}{green!10!orange!90!}

\newcommand{\ztbabs}{\texttt{ztbabs}}
\newcommand{\mekal}{\texttt{mekal}}
\newcommand{\pow}{\texttt{powerlaw}}
\newcommand{\fermi}{\textit{Fermi-LAT}}
\newcommand{\swift}{\textit{Swift}}
\newcommand{\xmm}{\textit{XMM-Newton}}

\received{17th July 2021}

\shorttitle{SED modeling of LLAGNs}
\shortauthors{Tomar et al.}
\graphicspath{{./}{figures/}}

\begin{document}
\title{Broadband Modeling of Low Luminosity Active Galactic Nuclei Detected in Gamma Rays}

\correspondingauthor{Gunjan Tomar}
\email{gunjan@rri.res.in}

\author[0000-0003-4992-6827]{Gunjan Tomar}
\affiliation{Raman Research Institute, Bangalore, India}

\author[0000-0002-1188-7503]{Nayantara Gupta}
\affiliation{Raman Research Institute, Bangalore, India}

\author[0000-0002-1173-7310]{Raj Prince}
\affiliation{Center for Theoretical Physics, Polish Academy of Sciences, Al.Lotnikov 32/46, PL-02-668 Warsaw, Poland}

\begin{abstract}
Low luminosity active galactic nuclei are more abundant and closer to us than the luminous ones but harder to explore as they are faint. We have selected the four sources NGC 315, NGC 4261, NGC 1275, and NGC 4486, which have been detected in gamma rays by \textit{Fermi-LAT}. We have compiled their long term radio, optical, X-ray data from different telescopes, analysed \textit{XMM-Newton} data for NGC 4486, \textit{XMM-Newton} and \textit{Swift} data for NGC 315. We have analysed the \textit{Fermi-LAT} data collected over the period of 2008 to 2020 for all of them. 
Electrons are assumed to be accelerated to relativistic energies in sub-parsec scale jets, which radiate by synchrotron and synchrotron self-Compton emission covering radio to gamma-ray energies. This model can fit most of the multi-wavelength data points of the four sources. However, the gamma-ray data points from NGC 315 and NGC 4261 can be well fitted only up to 1.6 GeV and 0.6 GeV, respectively in this model. This motivates us to find out the origin of the higher energy $\gamma$-rays detected from these sources. Kilo-parsec scale jets have been observed previously from these sources in radio and X-ray frequencies. If we assume $\gamma$-rays are also produced in kilo-parsec scale jets of these sources from inverse Compton scattering of starlight photons by ultra-relativistic electrons, then it is possible to fit the gamma-ray data at higher energies. Our result also suggests that strong host galaxy emission is required to produce GeV radiation from kilo-parsec scale jets.
\end{abstract}

\section{Introduction} \label{sec:intro}

 With almost complete unanimity, it is believed that most of the giant galaxies host supermassive black holes (SMBHs) at their centers (\citealt{magorrian}, \citealt{ferrarese}, \citealt{korm13}). Accretion onto these SMBHs powers the most persistent sources of electromagnetic radiation in the Universe known as Active Galactic Nuclei (AGNs). In the present day Universe, majority of the AGNs host underfed SMBHs which are accreting at low, sub-Eddington rates (with Eddington ratio, L$_{\rm bol}$/L$_{\rm Edd} \leq 10^{-3}$) as revealed by optical spectroscopic surveys (\citealt{ho97}, \citealt{ho08}). With average bolometric luminosity less than 10$^{42}$ erg s$^{-1}$ (\citealt{Terashima_2000}), these low luminosity AGNs (LLAGNs) occupy the fainter end of the AGN luminosity function. The low luminosity of these sources makes them incapable of sustaining structural features like broad line region (BLR; \citealt{blr}) and dusty torus (\citealt{torus}), which are cornerstones of the inclination-based unified scheme of AGNs (\citealt{ant}). 
 \par
Many LLAGNs can be described by advection-dominated accretion flows (ADAFs) where the plasma thermal energy is advected all the way into the event horizon before being radiated away. At sub-Eddington accretion rates, ADAFs are radiatively inefficient with low densities and low optical depth. This results in a geometrically thick and optically thin accretion flow unlike geometrically thin, optically thick accretion flows in luminous AGNs (see \citealt{narayan}). At high-mass accretion rate, the optical depth of accretion flow becomes high and most of the internal energy carried by photons, get trapped inside the flowing matter and reduces the radiative efficiency. This model is called optically thick ADAF, or `slim disk' (\citealt{abram}). 
We usually refer to the optically thin ADAF as radiatively inefficient accretion flow (RIAF). In this paper, we have used the terms ADAF and RIAF interchangeably, considering both represent the optically thin, geometrically thick accretion flows.
Observationally, it has been seen that the big blue bump which is a telltale signature for the standard accretion disk in more luminous AGNs, are either absent or weak in the spectral energy distributions of LLAGNs (\citealt{ho08}). Further, the conspicuous presence of red bumps in mid-IR band as well as the presence of double-peaked Balmer emission lines indicate the presence of optically thick outer truncated disk (\citealt{quat}, \citealt{ho20}). These suggest that the central engines go through fundamental changes as the accretion rate decreases to sub-Eddington limits, thus nullifying the hypothesis that the LLAGNs are the scaled-down versions of their more luminous predecessors.
\par

Observational and theoretical studies suggest that RIAFs are quite efficient at producing powerful bipolar outflows and jets owing to their vertical thick structure which enhances the large-scale poloidal component of magnetic field, which is crucial for the formation of jets (\citealt{narayan}, \citealt{nemmen07}, \citealt{nar_mc}). 
The radio cores in LLAGNs, detected using 15GHz VLA images by \citet{nagar05} indicates the primary accretion energy output is in jet kinetic power. 
Despite of their low luminosities, these objects are radio-loud (radio-loudness is anti-correlated with Eddington ratio; \citealt{ho02}) which further supports the existence of jets in these systems. 
Extended structures in radio are also seen from some LLAGNs when observed with sufficient angular resolution and sensitivity (\citealt{mez}).
\par
Perforce, it has been suggested that in most LLAGNs, emission comes from three components: a jet, RIAF and an outer thin disk \footnote{Thin disk, here, refers to the geometrically thin, optically thick disk.} (\citealt{nemmen}). Thus the jet, RIAF and RIAF with truncated thin disk models (e.g. \citealt{falc}, \citealt{merloni} and \citealt{yu11}) are widely used to explain the spectral energy distributions of LLAGNs.
Their relative contributions are not yet known.
\citet{nemmen10} modelled 24 LINERs to demonstrate that both ADAF-dominated and jet-dominated models can explain the observed X-ray data consistently. 
\par
  The $\gamma$-rays emitted from the non-thermal leptons that are accelerated in the jet can help to probe the jet component directly. This accelerated population of electrons in the magnetized jet emits synchrotron radiation, which is then up-scattered by the same electron population through inverse-Compton scattering producing $\gamma$-rays (\citealt{maras}). The modelling of broadband SED extending from radio to $\gamma$-rays helps to
  constrain the physical parameters of the jet.
\par
The Large Area Telescope instrument on board Fermi satellite (\fermi~) is a pair conversion $\gamma$-ray telescope covering an energy range from $\sim$ 20 MeV to more than 300 GeV (\citealt{lat}). It primarily operates in an all-sky survey mode, where it scans the entire sky in approximately every 3 hours. It has been surveying the entire sky in the energy range of 100 MeV to 300 GeV for more than 12 years (\citealt{atwood}).
\citet{ho95} conducted Palomar spectroscopic survey of northern galaxies selected based on apparent blue magnitude B$_T<$ 12.5 and found out that over 40\% of nearby galaxies contain LLAGNs. The Palomar survey is ideal for the study of demographics and physical properties of nearby galaxies, especially LLAGN since the survey is composed of high-quality, moderate-resolution, long-slit optical spectra (\citealt{ho08}). 
\citet{rani_llagn} confirmed that the four LLAGNs (NGC 315, NGC 4261, NGC 1275, NGC 4486) from the Palomar survey are $\gamma$-rays emitters with more than 5$\sigma$ significance, by analyzing 10.25 years of \textit{Fermi-LAT} data. 
To the best of our knowledge these are the only LLAGNs from the Palomar survey, which have been detected in $\gamma$-rays by \textit{Fermi-LAT}.
While NGC 1275 and NGC 4486 have been identified as $\gamma$-ray emitters before, NGC 315 and NGC 4261 are identified as gamma-ray emitters for the first time by \cite{rani_llagn}. They have shown that single-zone synchrotron self-Compton (SSC) emission from the jet  can explain the gamma-ray emission up to a few GeV for NGC 315 and NGC 4261 while hadronic emission from RIAF fails to do so. 
\par
In this work, we consider the four LLAGNs mentioned above from the Palomar survey that are detected in $\gamma-$rays, as our sample. We model their multi-wavelength spectral energy distributions with leptonic model assuming the emission regions are located in sub-parsec and kilo-parsec scale jets.
At kilo-parsec scales, it has been suggested that starlight from the galaxy can be a dominant photon field for inverse Compton scattering off electrons (\citealt{stawarz03}). This allows us to include a gamma-ray emission component from the kilo-parsec scale jet to explain the emission above a few GeV, which could not be explained by a single zone SSC emission from the sub-parsec scale jet. We have calculated the $\gamma$-ray emission from the kilo-parsec scale jet, produced by external Compton (EC) scattering of galactic starlight photons by the relativistic electrons in the kilo-parsec scale jet.
 The broadband spectral energy distributions are modelled using a time dependent code which includes radiative cooling and escape of relativistic leptons from the emission region. 
 \par 
 We have shown that SSC emission from the sub-parsec scale jet can well explain the broadband spectral energy distributions of these four LLAGNs, while external Compton emission from the kilo-parsec jet is required to explain the $\gamma$-ray emission beyond 1.6 GeV and 0.6 GeV in the case of NGC 315 and NGC 4261, respectively. \\
A standard $\Lambda$CDM cosmology model with H$_0$ = 75 km/Mpc/s and $\Omega_{matter} = $0.27 is assumed, throughout this paper.

In section \ref{sec:2}, we have discussed about the LLAGNs studied in this work.
The data analysis is discussed in section \ref{sec:3}. The modeling of spectral energy distributions and results are discussed in section \ref{sec:4}. The summary and conclusions are presented in Section \ref{sec:5}.

\section{Sample}
\label{sec:2}
\subsection{NGC 315}
NGC 315 is a nearby elliptical galaxy, located at a redshift of 0.01648 (\citealt{trager}). It hosts a Fanaroff-Riley type 1 (FR I) radio source with two-sided asymmetric well resolved radio jets at arcsec and milliarcsec resolutions shown both with Very Long Baseline Interferometry (VLBI) and Very Large Array (VLA) observations (\citealt{venturi}, \citealt{cotton}). The high spatial resolution of Chandra imaging allowed the detection of X-ray emission from the main jet (\citealt{wor}) inclined at an angle 38$^\circ \pm 2^\circ$ to our line of sight (\citealt{canvin}). The Hubble Space Telescope (HST) image shows a clear circum-nuclear dusty disk with 2.5\arcsec~ diameter in its center (\citealt{verdoes99}). It has been classified as a LLAGN by \citet{ho97} through the detection of broad H$_\alpha$ line. Its LLAGN nature was later confirmed by \citet{gu07}, who obtained the bolometric luminosity of L$_{bol} \sim 1.9 \times 10^{43}$ ergs s$^{-1}$ corresponding to an extremely low Eddington ratio of 4.9 $\times 10^{-4}$. 
\\
Located at a distance of 65.8 Mpc (\citealt{nagar05}) this LLAGN is one of the four low accreting galaxies from the Palomar survey (\citealt{ho95}, \citealt{ho97}) to be detected at $\gamma$-ray energies at above 5$\sigma$ significance by \fermi~(\citealt{rani_llagn}). It is detected with a statistical significance of $\sim9\sigma$ in the energy range 0.1-300 GeV over an observation of 10.25 years ranging from August 4th 2008 to November 15th 2018. The measured differential spectrum is well defined by a power-law with photon index $\Gamma$ = 2.32 $\pm$ 0.11 with an average flux of 3.38 $(\pm 0.43)\times10^{-12}$ ergs cm$^{-2}$ s$^{-1}$.

\subsection{NGC 4261} \label{sec:4261}
NGC 4261 is an elliptical galaxy located at a redshift of 0.00738 (\citealt{cap11}),  with a nucleus classified as a type 2 Low-ionization nuclear emission-line region (LINER) based on high-quality optical spectra by \citet{ho97}. It is located at a distance of 35.1 Mpc (\citealt{nagar05}) with a SMBH of mass 4.9 ($\pm$0.1)$\times 10^8 M_\odot$ (\citealt{ferrar}) at the center. It hosts a low-power FR I radio source with twin jets (\citealt{birkin}) oriented at angle 63$^\circ \pm3^\circ$ with respect to the line of sight of the observer (\citealt{piner}). In addition, a 300 pc-scale nuclear disk of gas and dust was imaged by HST (\citealt{jaffe}, \citealt{ferrar}) lying orthogonal to the radio jets. \\
The presence of X-ray jet has been detected in the inner few kpc-scale of the radio jets by Chandra observations (\citealt{gliozzi}, \citealt{zez05}, \citealt{worrall4261}). \citet{zez05} also showed a substantial absorbing column in X-rays. The luminosity after absorption and bolometric corrections is only 2.0$\times 10^{-5}$ of the Eddington luminosity, implying a low accretion rate.\\
NGC 4261 was detected in $\gamma$-rays by \fermi~ with a significance of $\sim 6.8\sigma$ over a period of 10.25 years (\citealt{rani_llagn}). A power-law with photon index 2.15 $\pm$ 0.16 and an average flux of 2.15 ($\pm$ 0.42)$\times 10^{-12}$ ergs cm$^{-2}$ s$^{-1}$ explains the measured spectral energy distribution in the energy range 0.1-300 GeV.

\begin{figure*}
    \centering
    \includegraphics[width=\textwidth]{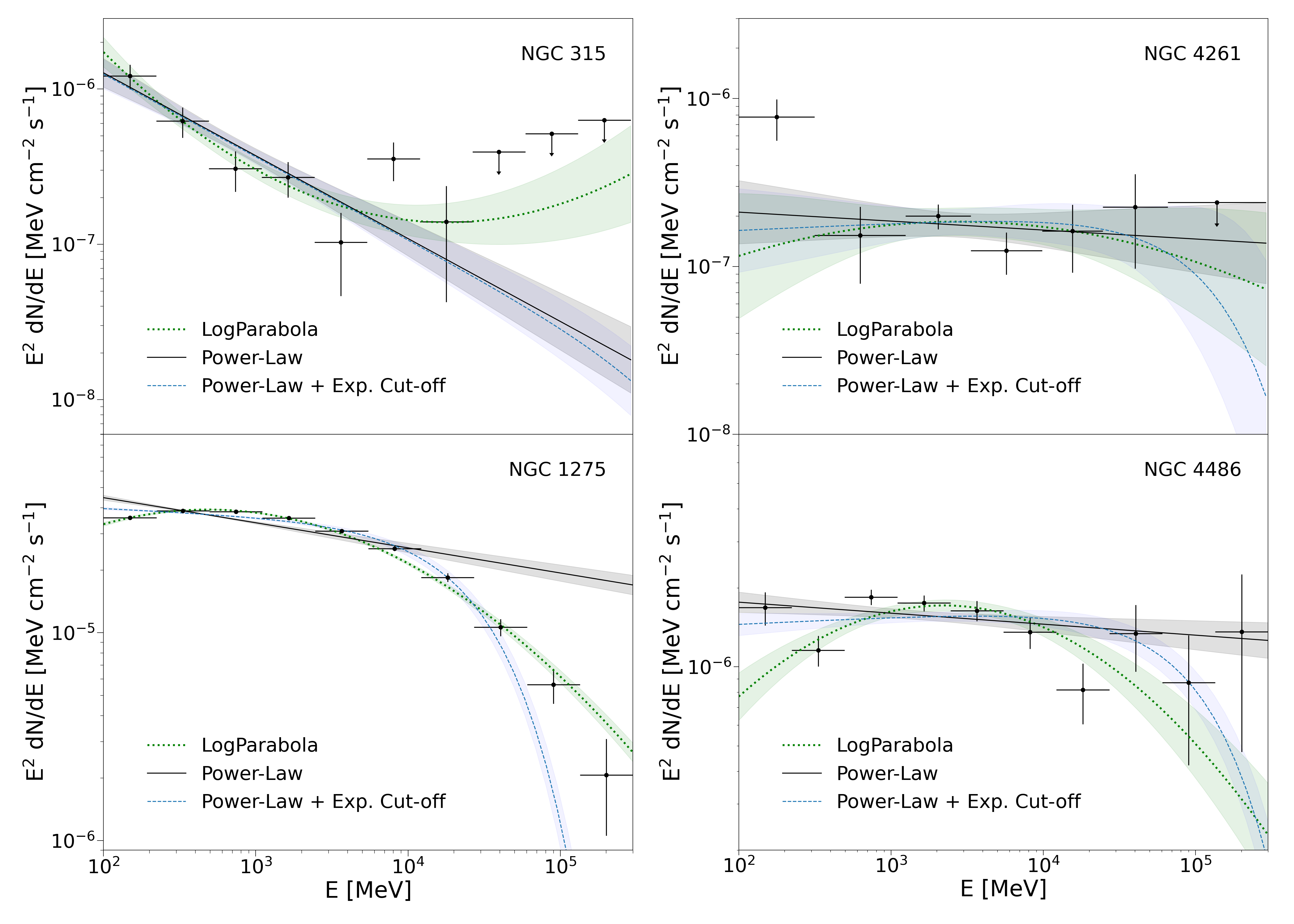}
    \caption{The $\gamma$-ray SED obtained by \textit{Fermi-LAT} from 12 years of observations. The shaded part corresponds to 1$\sigma$-uncertainty in the respective fit. When TS$<$4, upper-limits at 95\% confidence level are plotted with down arrow.}
    \label{fig:fer}
\end{figure*}

\begin{table*}
\centering
\begin{threeparttable}[b]
    \setlength{\tabcolsep}{0pt}
   \caption{Parameter results for $\gamma$-ray SED obtained by \textit{Fermi-LAT} fitted using different spectral models in the range 100 MeV to 300 GeV}
   \label{fermi_sed}
 \centering
     \begin{tabular*}{\textwidth}{@{\extracolsep{\fill}}lcccc@{}}
     
\hline
        \multirow{2}{*}{Source} & \multirow{2}{*}{Model} & \multirow{2}{*}{Parameter Values\tnote{a}} & Flux$_{0.1-300 GeV}$ & \multirow{2}{*}{TS$_{curve}$\tnote{b}} \bigstrut\\
        & & & \small{(ergs cm$^{-2}$ s$^{-1}$)} & \\
        \hline
        \hline
    \multirow{5}{*}{NGC 315 (4FGL J0057.7+3023)} &Power-Law &$\Gamma=2.53\pm0.11$ & 3.81$(\pm0.41)\times 10^{-12}$  & -  \bigstrut \\
        \cline{2-5}
        & \multirow{2}{*}{Log-Parabola} &$\alpha=2.52 \pm 0.08$ & \multirow{2}{*}{4.51$(\pm0.75)\times 10^{-12}$}  & \multirow{2}{*}{3.16} \bigstrut \\
         & & $\beta=-0.09\pm 0.03$ &  & \\
        \cline{2-5}
         & Power-Law with & $\Gamma=2.53\pm0.11$  & \multirow{2}{*}{3.73$(\pm0.40)\times 10^{-12}$} & \multirow{2}{*}{-1.7} \bigstrut \\
         & Exponential Cutoff & $E_{c}=0.9$ TeV &  & \\
        \hline
        \multirow{5}{*}{NGC 4261 (4FGL J1219.6+0550)} &Power-Law &$\Gamma=2.05\pm0.15$ & 2.22$(\pm 0.42)\times 10^{-12}$ & -  \bigstrut \\
        \cline{2-5}       
        & \multirow{2}{*}{Log-Parabola} &$\alpha=1.99\pm(0.19)$ & \multirow{2}{*}{1.92$(\pm0.69)\times 10^{-12}$}  & \multirow{2}{*}{-0.58} \bigstrut \\
         & & $\beta=0.04\pm0.09$ & & \\
        \cline{2-5}
         & Power-Law with & $\Gamma=1.95\pm0.22$  & \multirow{2}{*}{1.97$(\pm0.54)\times 10^{-12}$}  & \multirow{2}{*}{2.34} \bigstrut \\
         & Exponential Cutoff & $E_{c}=0.1$ TeV & & \\
         \hline
         \multirow{5}{*}{NGC 1275 (4FGL J0319.8+4130)} & Power-Law &$\Gamma=2.12\pm0.02$ & 3.70$(\pm0.05)\times 10^{-10}$ & -  \bigstrut \\
        \cline{2-5}
        & \multirow{2}{*}{Log-Parabola} &$\alpha=2.08\pm0.01$ & \multirow{2}{*}{3.10$(\pm 0.03)\times 10^{-10}$} & \multirow{2}{*}{383.6} \bigstrut \\
         & & $\beta=0.065\pm0.003$ & & \\
        \cline{2-5}
         & Power-Law with & $\Gamma=2.03\pm0.01$  & \multirow{2}{*}{3.02$(\pm 0.02)\times 10^{-10}$} & \multirow{2}{*}{351.1} \bigstrut \\
         & Exponential Cutoff & $E_{c}=30.8\pm2.6$ GeV & & \\
       \hline
     \multirow{5}{*}{NGC 4486 (4FGL J1230.8+1223)} & Power-Law &$\Gamma=2.04 \pm0.03$ & 1.94$(\pm 0.09)\times 10^{-11}$ & -  \bigstrut \\
        \cline{2-5}
        
       & \multirow{2}{*}{Log-Parabola} &$\alpha=1.94\pm0.04$ & \multirow{2}{*}{1.48$(\pm 0.11)\times 10^{-11}$} & \multirow{2}{*}{-18.6} \bigstrut \\
         & & $\beta=0.08\pm0.02$ & & \\
        \cline{2-5}
        
         & Power-Law with& $\Gamma=1.97 \pm 0.03$  & \multirow{2}{*}{1.69$(\pm 0.08)\times 10^{-11}$}  & \multirow{2}{*}{-51.1} \bigstrut \\
         & Exponential Cutoff & $E_{c}=0.13$ TeV & & \\
        \hline
\end{tabular*}
  
\begin{tablenotes}
    \item[a] Symbols are as defined in Section
    \item[b] TS$_{curve}$ = 2 (log$\mathscr{L} (\mathscr{M})$ - log$\mathscr{L}$(PL)) where $\mathscr{L}$ is the maximum likelihood of the model, $\mathscr{M}$ is either a Log-Parabola or a Power Law with Exponential Cutoff spectral model. PL is a Power-Law spectral model.
    \end{tablenotes}
 \end{threeparttable} 
\end{table*}

\subsection{NGC 1275}
NGC 1275 is one of the nearest radio galaxies at a redshift of z $=$ 0.0176 (\citealt{young}). It is located at a distance of 70.1 Mpc (\citealt{nagar05}) and is elliptical in shape. It is a radio-loud AGN with a relatively low Eddington ratio of 3$\times10^{-4}$ (\citealt{sikora}), classifying it as a low luminosity AGN. It is identified as Seyfert 1.5 due to the presence of weak broad emission line based on H$\alpha$ study (\citealt{ho97}). The detailed studies of the AGN with VLBI and VLA established the presence of an exceptionally bright radio source (3C 84) with asymmetrical jets at both parsec and kilo-parsec scales(\citealt{walker00}, \citealt{verm}, \citealt{asada}), suggesting an FR I morphology. These studies reveal a jet angle of 30$^\circ$-60$^\circ$ with our line of sight (\citealt{walker94}, \citealt{asada}). Recently, \citet{fujita} suggested the viewing angle of the jet with respect to the line of sight to be 65$^\circ \pm 15^\circ$ based on the increased radio activity detected by \citet{nagai10}.
\par
NGC 1275 is one of the brightest radio galaxies detected at the high energy (HE; $>$100 MeV) and very high energy (VHE; $>$100 GeV) $\gamma$-rays (\citealt{abdo1275}, \citealt{alek12}).
The average flux and the photon index measured by \fermi~ from 2008 August 4 to 2016 November 15 are F$_{>100MeV} = 3.34 (\pm 0.03 ) \times   10^{-7}$  ph cm$^{-2}$ s$^{-1}$ and 1.93 $\pm$ 0.01, respectively (\citealt{tanada18}).
At VHE, \textit{MAGIC} measured the average $\gamma$-ray flux above 100 GeV to be 1.3 ($\pm$ 0.2) $\times$ $10^{-11}$ph cm$^{-2}$ s$^{-1}$. The corresponding differential spectrum in 70-500 GeV was estimated with a power-law of photon index 4.1 $\pm$ 0.7 (\citealt{alek12}). This spectral break from HE to VHE was later confirmed by several studies like \citet{fuka18}, \citet{tanada18}.
 NGC 1275 has also been observed at VHE by other imaging atmospheric Cherenkov telescopes like \textit{HEGRA, Whipple, VERITAS} (\citealt{mukh}) and \textit{TACTIC} (\citealt{gho20}).
\subsection{NGC 4486 (M87)}
M87 is a giant elliptical galaxy located in Virgo cluster at a redshift, $z=0.00428$ (\citealt{cap11}). It is located at a distance of 16.8 Mpc (\citealt{nagar05}), with a SMBH of mass 6.5 $\times 10^9 M_{\odot}$at its center powering it (\citealt{eht}). Despite of hosting such a SMBH, its bolometric luminosity is only of the order of 10$^{42}$ ergs s$^{-1}$, which is six orders of magnitude lower than the Eddington luminosity (\citealt{reynolds}), placing this in the class of LLAGNs. It is commonly classified as a FR I radio galaxy. The presence of narrow emission lines also suggests its a type 2 LINER. 
The relativistic jet, first detected by H. Curtis (\citealt{curtis}) in optical is misaligned with respect to the line of sight with an angle between 15$^\circ$ and 30$^\circ$ (\citealt{biretta}; \citealt{acciari}; \citealt{walker}). Due to its proximity, the jet is well-imaged at radio through X-ray frequencies. This relativistic outflow extends upto kilo-parsec scales (\citealt{marshall02}) and its radiative output is believed to dominate the spectral energy distribution of the AGN core (\citealt{abdo09}, \citealt{nemmen}, \citealt{dejong}, \citealt{prieto}, \citealt{fraija}).
Recent polarised image of the SMBH of M87 indicates the presence of strong magnetic field at the event horizon which can launch powerful jets (\citealt{eht21}).
\par
M87 is the first extra-galactic object to be detected at VHE by \textit{HEGRA} (\citealt{ah2003}). Since then, it has been detected at $\gamma-$ray frequencies by \textit{H.E.S.S., VERITAS, MAGIC,} and \fermi~(\citealt{ah06}, \citealt{al08}, \citealt{ac08}, \citealt{abdo09}).

\section{Multiwavelength observations and Data Analysis}
\label{sec:3}
We have constructed the multiwavelength SED of our sample by compiling radio to UV data on these sources from earlier works.  
We have analysed the X-ray data recorded by \textit{XMM-Newton} in 2017 from NGC 4486, \textit{XMM-Newton} data taken in 2019 and \textit{Swift} data from 2017 to 2018 for NGC 315, and compiled the archival X-ray data available on NGC 4261 and NGC 1275 (see details in Section \ref{sec:xray}). In addition, we analysed 12 years of \fermi~data, collected over the period of 2008 to 2020. Due to low spatial resolution in gamma-ray energy band, the distinction between the emission from the sub-parsec scale and kilo-parsec scale jet cannot be made.  The radio and X-ray data points of the extended jets of NGC 315 and NGC 4261 were obtained from observations by \textit{Very Large Array (VLA)} and \textit{Chandra} observatories respectively owing to their high spatial resolution (\citealt{worrall315}, \citealt{worrall4261}).
The results obtained for the maximum likelihood analysis of \fermi~ data are summarized in Table \ref{fermi_sed} for different spectral models. The 12 years averaged spectrum along with the best-fit of each spectral model are shown in Figure \ref{fig:fer}. Data from NED \footnote{\url{https://ned.ipac.caltech.edu/}} has also been taken for an overall reference SED.

\subsection{Radio to UV}
\begin{figure*}
    \centering
    \includegraphics[width=\textwidth]{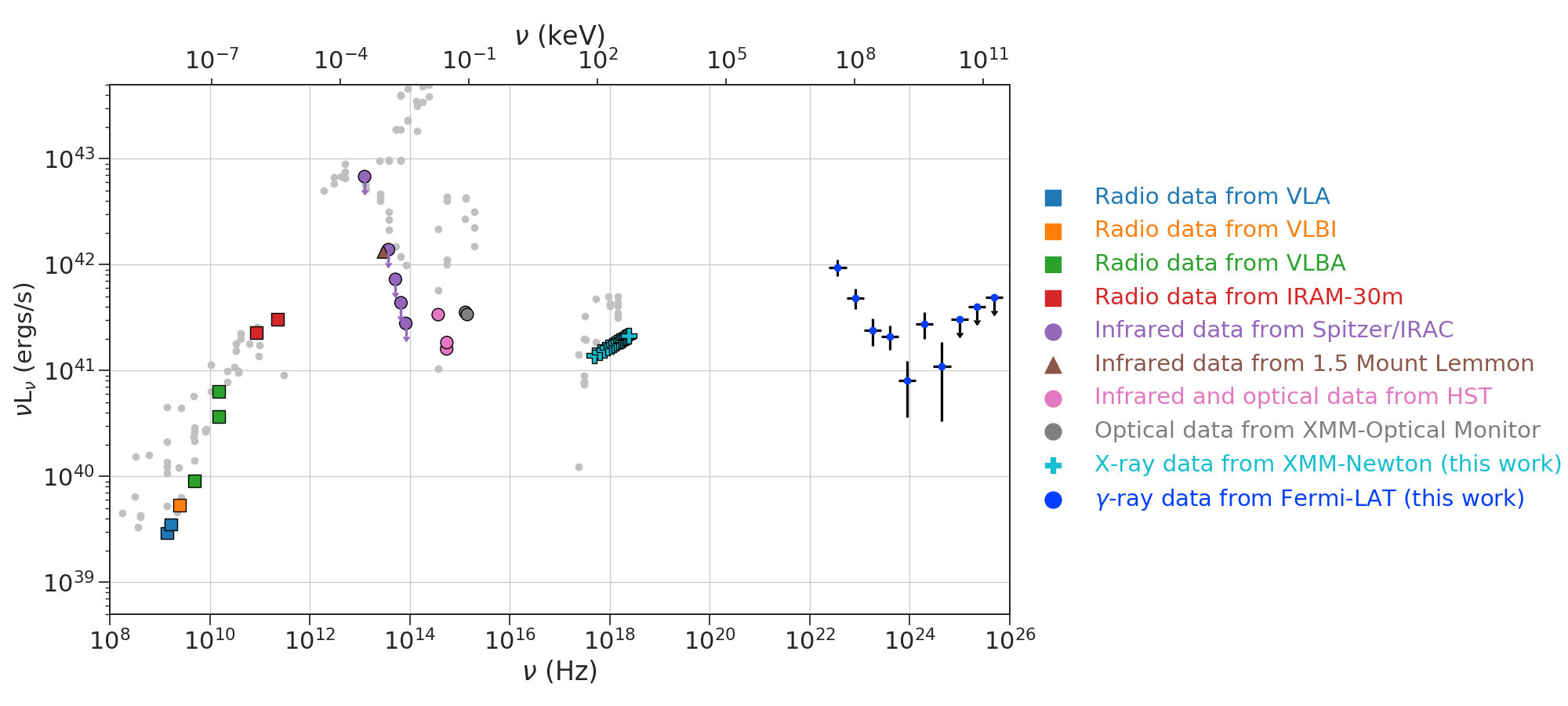}
    \caption{The multiwavelength SED of NGC 315 constructed using radio data taken from \citet{cap05}, \citet{lazio}, \citet{nagar05}, \citet{agudo}, \citet{kov} and \citet{venturi}, infrared data taken from \citet{gu07}, \citet{heck} and \citet{verdoes}, optical data taken from \citet{verdoes} and \citet{younes}, X-ray data taken from \citet{worall}. Other data points from NED are shown in silver. }
    \label{fig:315}
\end{figure*}

\begin{figure*}
    \centering
    \includegraphics[width=\textwidth]{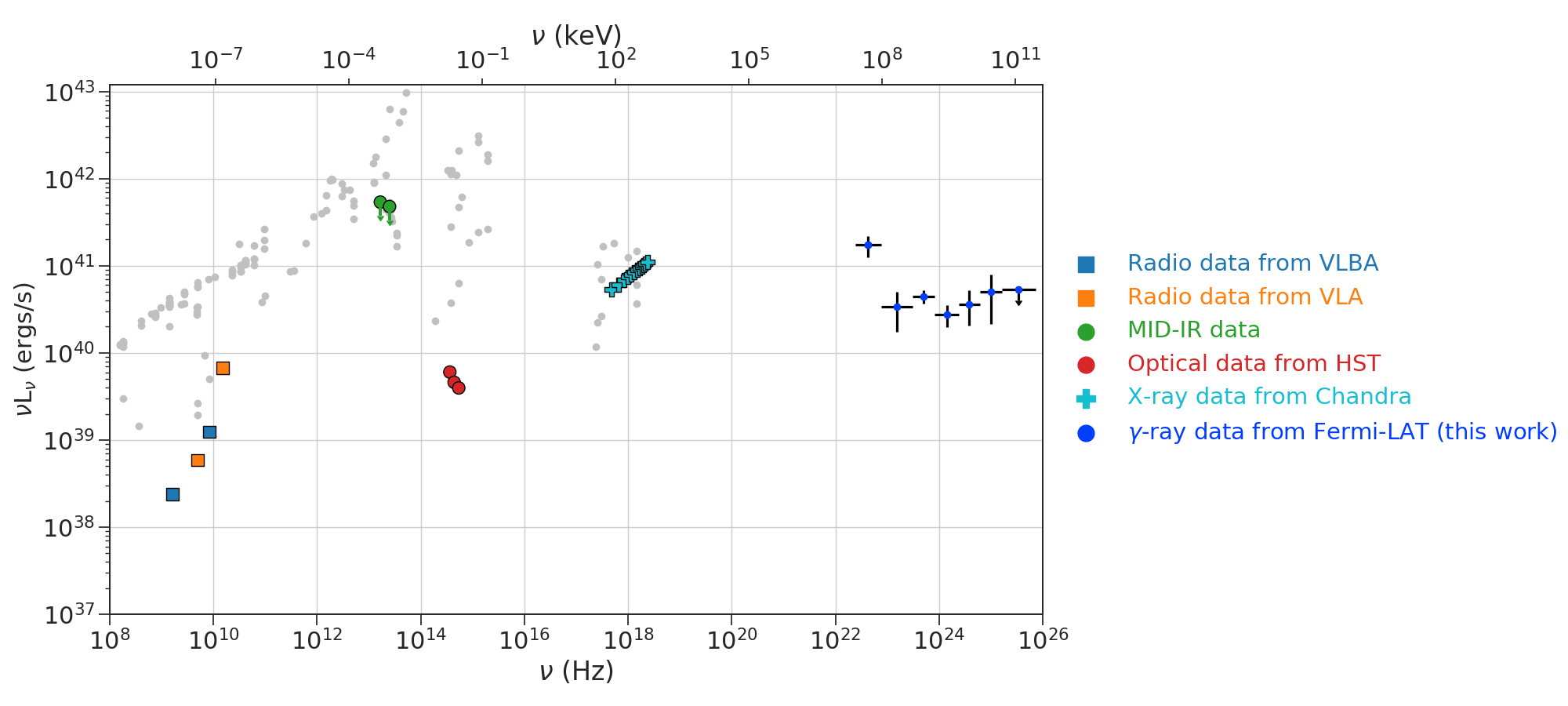}
    \caption{The multiwavelength SED of NGC 4261 constructed using radio data taken from \citet{jones}, \citet{nagar05}, infrared data taken from \citet{asmus}, optical data taken from \citet{ferrar}. The X-ray observation is taken from \citet{zez05}. Other data points from NED are shown in silver. The higher flux in radio are low resolution measurements which could have significant contribution from the radio lobes of NGC 4261. }
    \label{fig:4261}
\end{figure*}

\begin{figure*}
    \centering
    \includegraphics[width=\textwidth]{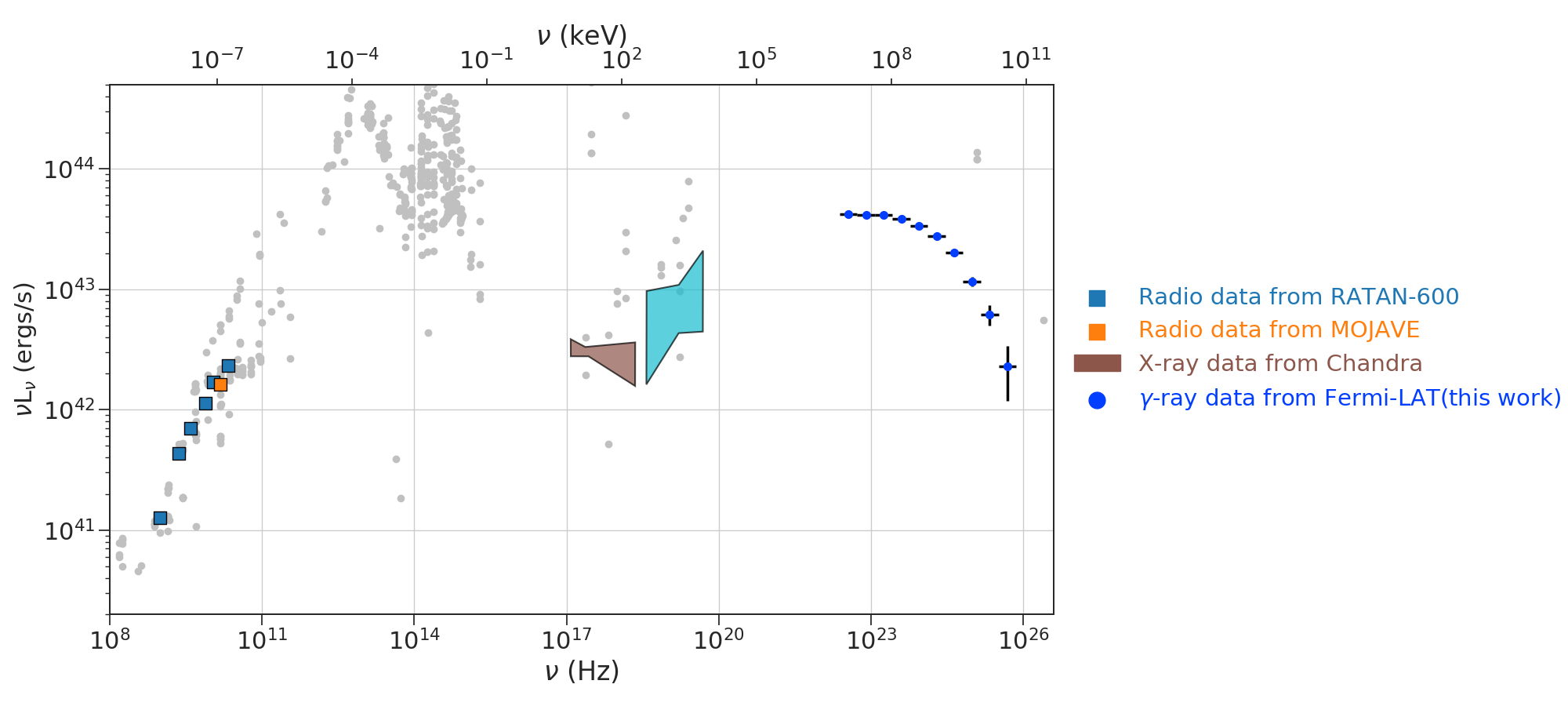}
    \caption{The multiwavelength SED of NGC 1275 is constructed using radio data from \citet{abdo1275}. X-ray data has been taken from \citet{tanada_chandra} and \citet{ajello}. Other data points from NED are shown in silver.}
    \label{fig:1275}
\end{figure*}

\begin{figure*}
    \centering
    \includegraphics[width=\textwidth]{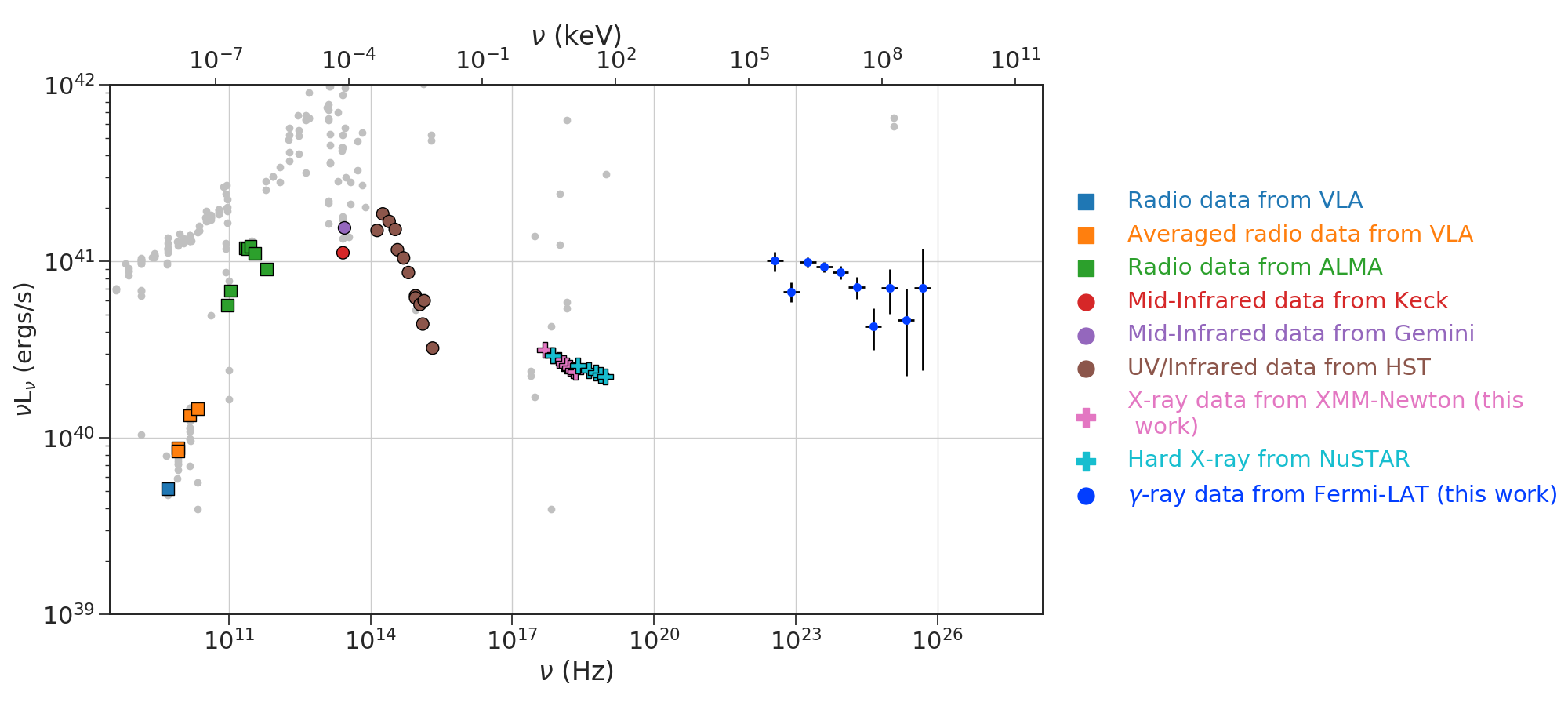}
    \caption{The multiwavelength SED of NGC 4486 constructed using data taken from \citet{nag1}, \citet{prieto}, \citet{why} and \citet{perl}. Other data points from NED are shown in silver. The higher flux in radio are low resolution measurements which could have significant contribution from the radio lobes of NGC 4486. }
    \label{fig:m87}
\end{figure*}

The radio to UV data for this work has been compiled from previous observations. The details of the observations and reductions can be found in the references given in Figure \ref{fig:315}, \ref{fig:4261}, \ref{fig:1275}, and \ref{fig:m87}. \\

Whenever available, radio data has been taken from NRAO\footnote{NRAO is a facility of the National Science Foundation operated under co-operative agreement by Associated Universities, Inc.} Very Large Array (VLA), NRAO Very Large Baseline Array (VLBA) and Very Large Baseline Interferometry (VLBI). The high resolution of these radio telescopes allowed to isolate radio emission from the AGN from the other sources. In optical band, data points from \emph{Hubble Space Telescope (HST)} has been obtained, if available.\\
\emph{NGC 315:} NGC 315 has been observed in radio band by several studies using VLA, VLBA and VLBI (\citealt{cap05}, \citealt{nagar05}, \citealt{kov}, \citealt{venturi}). As part of polarimetric survey, it was simultaneously observed at 86 GHz and 229 GHz in August 2010 using the XPOL polarimeter on the IRAM 30m radio telescope (\citealt{agudo}). 
 Infrared data at arcsec resolution has been obtained from \emph{Spitzer}. These are considered as upper-limits due to non-negligible contribution from the central dusty disk of the AGN.
The data points obtained using filter UVW2 and UVM2 of XMM-Optical monitor, in July 2005 by \citet{younes} are also included. For the modeling, HST data points are preferred over XMM-optical monitor data points due to lower resolution of the latter.\\
\emph{NGC 4261:} We have used the radio to UV data as compiled by \citet{rani_llagn}. The radio data from VLA and VLBI has been taken (\citealt{jones}, \citealt{nagar05}). The mid-infrared data was taken as sub-arcsec resolution images obtained using VISIR (\citealt{asmus}). \\
\emph{NGC 1275: } Almost simultaneous radio observations taken in August - September 2008 by VLBA as part of Monitoring Of Jets in Active galactic nuclei with VLBA Experiments (MOJAVE) and 600 meter ring radio telescope RATAN-600 of the  Special  Astrophysical  Observatory,  Russian  Academy of  Sciences, have been used in this work (\citealt{abdo1275}).\\
\emph{NGC 4486:} The radio to UV data compiled for the quiscent phase from aperture radius $\sim$ 0.4 arcsec by \citealt{prieto} has been used in our study. \\
The radio emission from the kilo-parsec jets of NGC 315 and NGC 4261 have been observed owing to the good resolution of VLA. The radio flux of 74 mJy and 50 mJy are observed at 5 GHz for the kilo-parsec jet [in the region between 3.2 to 16.2 arcsec from the nucleus] of NGC 315 (\citealt{worrall315}) and [measured in the region between 8.8 to 31.7 arcsec from the nucleus] of NGC 4261 (\citealt{worrall4261}), respectively. 
\subsection{X-ray}
\label{sec:xray}
\emph{NGC 315: }  Due to low spatial resolution, jet emission can not be distinguished from the core\footnote{Here, we define core as the data corresponding to the unresolved structures} in \textit{XMM-Newton} and \textit{Swift} observations. Recent observation of 2019 taken by \xmm~ and multiple observations taken by \swift~ between 2017 and 2018 were combined and analysed. The details on the data extraction and subsequent analysis for each instrument are given in Section \ref{xmm_an} and \ref{swift}.
\par
Owing to the high spatial resolution of \textit{Chandra} observatory, X-ray jet emission has been resolved between 3.6 and 16.2 arcsec from the core. The power-law index for X-ray spectrum is calculated to be $\alpha=1.2 \pm 0.2$ with an X-ray luminosity of 4.3 ($\pm 0.2$) $\times$ 10$^{40}$ ergs s$^{-1}$ in 0.3-5.0 keV (\citealt{worrall315}).
\\
\emph{NGC 4261: } Due to faintness, low counts were recorded by recent \swift~ observations and did not provide a good data set and thus, we compile the previous measurement in X-ray for NGC 4261. The soft X-ray data was recorded by \emph{Chandra} with an exposure time of ~35 ks. The flux (2 - 10 keV) and photon index are reported as 6.97$\times 10^{-13}$ ergs cm$^{-2}$ s$^{-1}$ and 1.54$^{+0.71}_{-0.39}$, respectively (\citealt{zez05}).
\par
\citet{worrall4261} analysed 100 ks \textit{Chandra} observation and could resolve jet emission out to 31.7 arcsec with a photon spectral index $\alpha = 1.22 \pm 0.22$ with an X-ray luminosity of 2.9 ($\pm 0.2$) $\times$ 10$^{39}$ ergs s$^{-1}$ in 0.3-5.0 keV.\\
\emph{NGC 1275: }  The most recent observations with \swift~ for NGC 1275, suffered from pile-up effect and were not analysed. The soft X-ray data taken from \emph{Chandra} and hard X-ray data taken from \emph{Swift-BAT} were reconstructed from the previous literatures. The photon index and the total integral flux in the 2–10 keV band reported, are 2.11 $\pm$ 0.16 and 1.14 $\times$ 10$^{-11}$ ergs cm$^{-2}$ s$^{-1}$ (\citealt{tanada18}). Though \emph{Swift-BAT} observations could not resolve the nucleus spatially, non-thermal hard X-ray emission from the nucleus with a photon index of 1.7$^{+0.3}_{-0.7}$ was inferred. The corresponding luminosity was reported as 8$ \times$ 10$^{42}$ ergs s$^{-1}$ in 0.5-8 keV energy band (\citealt{ajello}).
\\
\emph{NGC 4486: } There are multiple observations with \textit{XMM-Newton} from 2017. We analyse the observation with the maximum exposure time. The details of the analysis are given in Section \ref{xmm_an}.
\\
The spectral analysis for the reduced data from \xmm~ and \swift~ has been performed using  XSPEC (\citealt{xspec}) version 12.10.0f. The errors are quoted at 90\% confidence level.

\subsubsection{XMM-Newton}
\label{xmm_an}
X-ray Multi-Mirror Mission (\xmm~; \citealt{jansen}) carries three co-aligned X-ray telescopes observing in an energy range 0.1 - 15 keV. We utilize the data from three European Photon Imaging Cameras (EPICs) onboard XMM: PN (\citealt{struder}), MOS1, and MOS2 (\citealt{turner}). The observation data files are obtained from the XMM-Newton Science archive\footnote{\url{http://nxsa.esac.esa.int/nxsa-web/home}} (XSA) and are reduced using the Science Analysis System (SAS; v18.0.0). The event list obtained is filtered of any high flaring background activity which are further used to obtain the science products using single events (\emph{PATTERN = 0}) $\ge$ 0.3 keV for MOS and upto 12 keV for EPIC-pn camera. Only events with pattern 0–4 (single and double) for the PN and 0–12 for the MOS were selected. The EPIC observations are corrected for pile-up, if present. The response matrix files using the SAS task \emph{rmfgen} is generated, while the ancillary response files are generated using the task \emph{arfgen} in SAS. 
\\
The spectral analysis was performed in the energy range 0.5-10.0 keV. The photo-electric cross-sections and the elemental abundances of \citet{wilms00} are used throughout to account for absorption by neutral gas. An absorbed Galactic column density fixed to the value obtained by the \texttt{w3Nh} tool\footnote{\url{https://heasarc.gsfc.nasa.gov/cgi-bin/Tools/w3nh/w3nh.pl}} was applied to the spectral models to account for the Galactic absorption.
\\
\textit{NGC 315: }An observation of $\sim$ 50.9 ks of NGC 315 was made by \xmm~ on 2019 January 27 (Obs ID: 0821871701). The EPIC-pn and MOS cameras on board \xmm~ were operated in Prime Full Frame using the medium filter. We extract the source events from a circle of radius 25 arcsec centered on the source. The background events are extracted from a circle of radius 50 arcsec on the same CCD away from source. We group the spectra to have a signal-to-noise ratio of 3 with a  minimum of 20 counts per bin to have a better statistics, thus allowing the use of the $\chi^2$ statistics. All the three EPIC spectra were fitted simultaneously allowing for a cross-normalisation between them. The spectra is fitted with model \texttt{(ztbabs*po)}$+$\mekal~ with a reduced-$\chi^2$ value 1.18 (190 d.o.f). The model \ztbabs~and \mekal~take into account the intrinsic absorption and the thermal emission below 2 keV, respectively. We obtain a power-law index of 1.73$\pm$0.10 and an average flux of 5.7$^{+0.3}_{-0.4} \times 10^{-13}$ ergs cm$^{-2}$ s$^{-1}$ over an energy range 2.0-10.0 keV.
\\
\textit{NGC 4486: }We analyse a 132 ks observation (Obs. ID:0803670501) taken on July 6th 2017. Following the analysis of XMM-Newton by \citet{bohringer}, we extracted source spectra from circular regions of 4\arcsec~for each PN, MOS1 and MOS2. The background spectra for each were extracted from circular region of 20\arcsec~away from the source. We group the spectra to have a signal-to-noise ratio of 3. For PN and MOS1, we group the spectra to have minimum 25 counts per bin. Due to low counts in MOS2, we bin the spectra with minimum 1 count per bin. Following this, the three spectra could not be fitted simultaneously, since $\chi^2$ statistics could not be used for MOS2 spectrum. We fit the PN spectrum with a \pow~ plus \mekal~model with reduced-$\chi^2$ value 1.21 (75 d.o.f.). The addition of another absorption component for intrinsic absorption was not significant. The flux and power-law index obtained for the best-fit in the energy range 2.0-10.0 keV are 1.51$^{+0.02}_{-0.13}\times 10^{-12}$ ergs cm$^{-2}$ s$^{-1}$ and 2.19$^{+0.22}_{-0.30}$ respectively. We obtain a temperature value of 1.37$^{+0.25}_{-0.23}$ keV with abundance set at 1, which is compatible with the values obtained by \citet{donato}. We use the same model for MOS1 and MOS2 spectra. A power-law index of 2.14$^{+0.19}_{-0.23}$ and corresponding flux of 1.70$^{0.16}_{-0.19}\times 10^{-12}$ ergs cm$^{-2}$ s$^{-1}$ in 2.0-10.0 keV range are obtained for MOS1 with reduced-$\chi^2$ value 1.3 (34 d.o.f.). Due to poor statistics, the parameters could not be well constrained for MOS2 and thus, are ignored (though a rough-fit provides values compatible with those obtained with PN and MOS1 spectra).

\subsubsection{Swift}
\label{swift}
\swift~ X-ray telescope (XRT; \citealt{burrows}) is a focusing X-ray telescope onboard Neil Gehrels \emph{Swift} Observatory, operating in the energy range of 0.3 - 10 keV. We have analysed archival \emph{Swift}-XRT data, observed in Photon-Counting (PC) mode.
XRT data reduction was performed using the standard data pipeline package (\texttt{XRTPIPELINE v0.13.5}) in order to produce cleaned event files. Source events are extracted within a circular region with a radius of 30 arcsec centered on the source positions, while background events are extracted from source-free region of radius 60 arcsec close to the source region of interest.
The spectra is obtained from the corresponding event files using the \texttt{XSELECT } v2.4g software; we created the ancillary response file using the task \texttt{xrtmkarf}. The photo-electric cross sections and the solar  abundances  of \citet{wilms00} are used to account for absorption by neutral gas. An absorbed Galactic column density  derived for the source from \citet{karl} (obtained with w3Nh tool) was applied to the spectral model.
\\
\textit{NGC 315: }Due to low counts in a single spectra, we combine the spectra at multiple epochs using the \texttt{FTOOLS} task \textit{addspec}. The background spectra are summed using task \texttt{mathpha}. We bin the spectra as to contain minimum of 5 count per bin using task \texttt{grppha}. We use Cash statistics instead of $\chi^2$ statistics since the number of counts per bin is lesser than 20. We summed eleven spectra obtained between 2017 and 2018 to get a total exposure of 20.8 ks. We fit the resultant spectrum using an absorbed power law \texttt{(ztbabs*po)} and found a good fit with C-stat of 76.12 for 52 d.o.f. The residuals below 2 keV are further modeled by adding a \texttt{mekal} component at kT = 0.56 keV, improving the C-stat with a value of 58.18 (for 50 d.o.f.). This value is in agreement with the value found by \citet{gonz06}. The average flux and photon index obtained in the energy range 2.0-10.0 keV are 3.7$^{+0.8}_{-1.6}\times10^{-13}$ ergs cm$^{-2}$ s$^{-1}$ and 2.11$^{+0.9}_{-0.6}$, respectively. 

\subsection{Gamma-ray Data from Fermi-LAT}
The data collected by \fermi~during a period of 10.25 years for NGC 315 and NGC 4261 was analysed by \citet{rani_llagn}. We analysed the data set collected over a period of 12 years ranging from 2008 August 4 to 2020 August 21 for all 4 sources with \texttt{fermitools} v2.0.0, \texttt{fermipy} v1.0.0 (\citealt{fermipy}), and Pass 8 event processed data (\citealt{pass8}). 
The events are selected in 100 MeV to 300 GeV energy range in a 15$^{\circ} \times$ 15$^{\circ}$ region of interest (ROI) centered on the positions of each AGN. The data are binned spatially with a scale of 0.1$^{\circ}$ per pixel and 8 logarithmically spaced bins per energy decade.
\\
We only selected the \texttt{Source} class events (\texttt{evclass = 128} and \texttt{evtype = 3}) with the recommended filter expression (\texttt{DATA\_QUAL$>$0} \&\& \texttt{LAT\_CONFIG $==$ 1}). Also, a maximum zenith angle cut of 90$^{\circ}$ was applied to reduce the contamination from secondary $\gamma$-rays from the Earth limb.
\\
 We included the standard diffuse templates, “\texttt{gll\_iem\_v07}" and “\texttt{iso\_P8R3\_SOURCE\_V2\_v1}”, available from the Fermi Science Support Center \footnote{\url{https://fermi.gsfc.nasa.gov/ssc/data/access/lat/BackgroundModels.html}} (FSSC), to model the Galactic diffuse emission and isotropic extragalactic emission, respectively.
 \\
 To quantify the significance of $\gamma$-ray detection from each source, we used the test-statistics  (TS \footnote{TS = -2 (log$\mathscr{L}_{o}$ - log$\mathscr{L}_{1}$) where $\mathscr{L}_{o}$, $\mathscr{L}_{1}$ are the maximum likelihood for the model without an additional source and with an additional source at the specified location, respectively \citealt{likeli}.}) obtained in binned likelihood analysis using \texttt{minuit}. 
 
\subsubsection{Spectral Models for Fitting Gamma-ray Data}
\label{sec:fermi}
A binned maximum likelihood analysis is performed by taking into account all the sources included in the updated fourth source catalog (4FGL-DR2; \citealt{4fgl-cat}, \citealt{4fgldr}) and lying up to 5$^{\circ}$ outside the ROI in order to obtain the spectral parameters and the significance of detection of the source.
\par
An automatic optimization of the ROI was performed using function \texttt{optimize} within the package to ensure that all the parameters are close to their global likelihood maxima. To look for any additional sources in our model which are not included in the 4FGL (or 4FGL-DR2) catalog, we used \texttt{find\_sources()} with a power-law model with index 2, \texttt{sqrt\_ts\_threshold = 5.0} and \texttt{min\_seperation = 0.5}. Additional sources when detected with TS$>$25 were included during the LAT analysis. 
\par
The normalisation of all the sources with a radius of 5$^{\circ}$ from the ROI and the isotropic and Galactic diffuse emission templates were left to vary. The spectral shape parameters of the four LLAGNs were also kept free while those of the other sources were fixed at the values in the 4FGL catalog.
\\

The following spectral models are explored for the whole energy range :
\begin{enumerate}
    \item Power-Law :
    \begin{equation}
        \frac{dN(E)}{dE} = N_o \left(\frac{E}{E_o}\right)^{-\Gamma}
    \end{equation}
    where the normalization $N_o$ and $\gamma$-ray photon index $\Gamma$ are considered as free parameters. The scale value $E_o$ is fixed at its catalog value (\citealt{4fgl}, \citealt{lott}).

    \item Log Parabola:
    \begin{equation}
        \frac{dN(E)}{dE} = N_o \left(\frac{E}{E_o}\right)^{-\alpha-\beta ln(E/E_o)}
    \end{equation}   
    where $N_o$, $\alpha$ and $\beta$ are the free parameters.
    \item Power Law with exponential cut-off:
        \begin{equation}
        \frac{dN(E)}{dE} = N_o \left(\frac{E}{E_o}\right)^{-\Gamma}exp\left(\frac{-E}{E_c}\right)
    \end{equation}
    where $N_o$, $\Gamma$ and $E_c$ are the free parameters. 
\end{enumerate}

\begin{figure*}
    \centering
    \includegraphics[width=\textwidth]{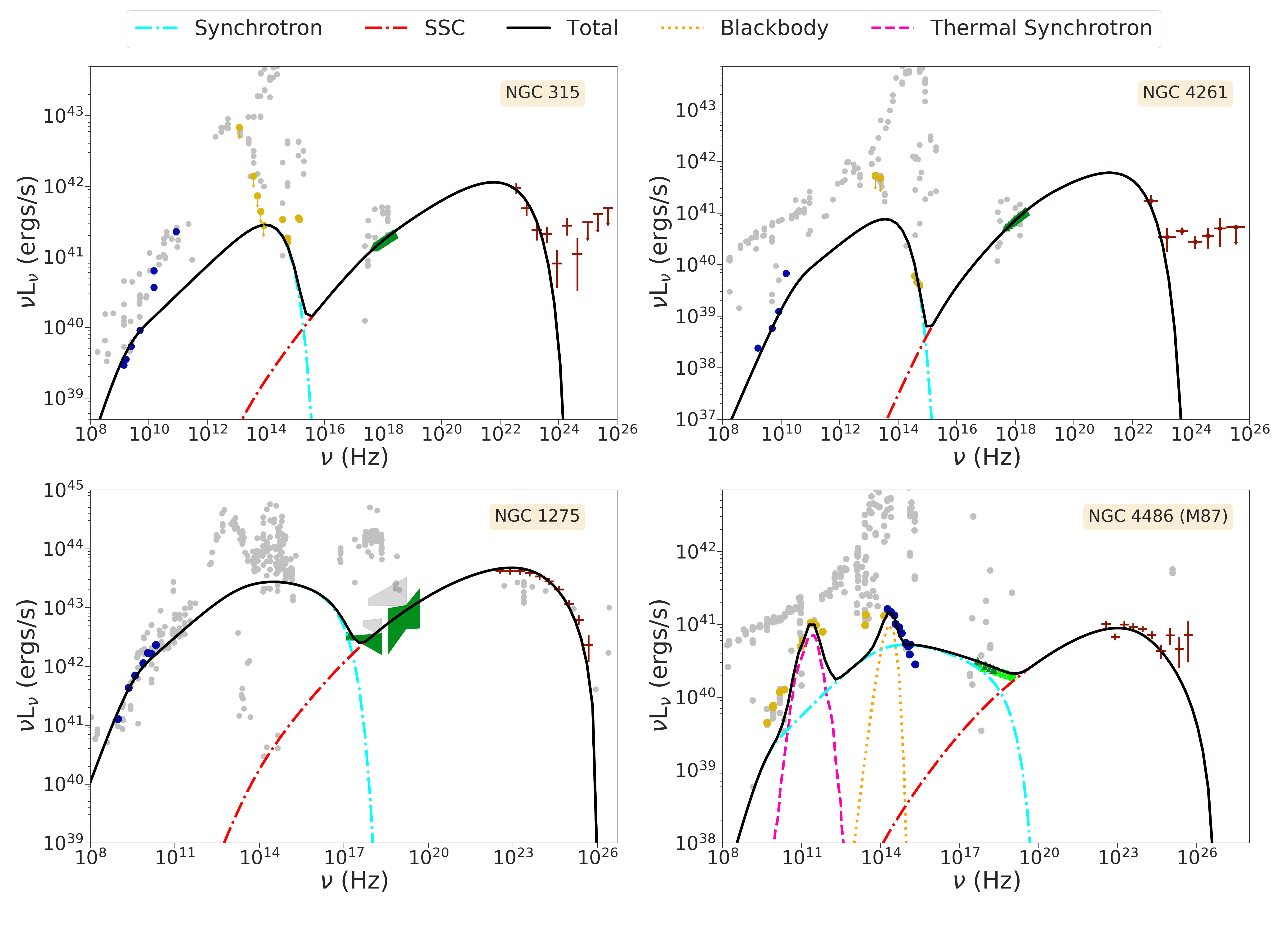}
    \caption{One zone leptonic modeling of broadband emission from sub-parsec scale jets of the four LLAGNs. The thermal synchrotron and the blackbody components of M87 have been reproduced from \citet{lucchini}. The gamma-ray data beyond 1.6 GeV from NGC 315 and 0.6 GeV from NGC 4261 cannot be explained with one zone SSC model.}
    \label{fig:sedf}
\end{figure*}
\subsubsection{Results of Gamma Ray Spectral Analysis}
To determine the best-fit spectral model of each source, the significance of spectral curvature is determined. The spectral curvature is significant if TS$_{curve}>$ 16 (corresponding to 4$\sigma$; \citealt{tsc}).
\\
 \emph{NGC 315: }No significant curvature is seen in the $\gamma$-ray SED of NGC 315. Its $\gamma$-ray spectrum is defined by a power-law with $\Gamma=2.53\pm0.11$ detected with $\approx$ 11.1 $\sigma$ (TS=123.7). An integrated flux of 3.1 ($\pm$ 0.41)$\times 10^{-12}$ ergs cm$^{-2}$ s$^{-1}$ over 0.1-300 GeV is obtained.
 \\
 \emph{NGC 4261: }It is the faintest in our sample with statistical significance detection of 8.2$\sigma$ (TS $=$ 67.24). As can be seen in Figure \ref{fig:fer}, no cut-off is seen in the spectrum. The 12 years averaged spectrum obtained is well represented by a power-law with $\Gamma=2.04\pm0.15$ with corresponding integrated flux over 0.1-300 GeV of 2.27$(\pm0.4)\times 10^{-12}$ ergs cm$^{-2}$ s$^{-1}$. 
  \\
 \emph{NGC 1275: }Figure \ref{fig:fer} clearly indicates a curvature in the spectrum. The source is detected with a high-statistical significance of 376.6$\sigma (TS=141829.6)$ with our likelihood analysis. The log-likelihood ratio test ($\Delta$TS$\sim$383.6 i.e. 19.6$\sigma$) signifies that a log-parabola model is a better representation over a single power-law model with an integrated flux of 3.1 ($\pm$0.03) $\times$ 10$^{-10}$ ergs cm$^{-2}$ s$^{-1}$ over the energy range 0.1-300 GeV with best-fit indices $\alpha=2.08\pm0.01$ and $\beta=0.065\pm0.003$. 
\\
\emph{NGC 4486: }The source was detected with high statistical significance TS$=$1844.52 ($\sim 43\sigma$) with our likelihood analysis.The average spectrum is well-defined by a power law of photon index, $\Gamma= 2.04\pm0.03$ and integrated flux, F$_{0.1-300 GeV}$ = 1.94 ($\pm$ 0.09) $\times$ 10$^{-11}$ ergs cm$^{-2}$ s$^{-1}$.

\section{Multi-wavelength SED Modeling}
\label{sec:4}
 We consider homogeneous and spherical emission region of radius $R$ moving through the magnetic field $B$ inside the jet with a bulk Lorentz factor $\Gamma_b$. This region contains relativistic plasma of electrons and protons and emit radiation through the synchrotron and inverse-Compton processes. 
\par
A simple power-law injection spectrum is expected in the case of Fermi-I type acceleration. While a power-law is supported by the $\gamma$-ray SED fits for NGC 315, NGC 4261 and NGC 4486, presence of curvature in the $\gamma$-ray SED of NGC 1275 hints at a different particle distribution. As suggested by \citet{massaro}, the injected particles may show an intrinsic curvature following a log-parabolic distribution  due to energy dependent acceleration, which is supported by the best-fit results for $\gamma$-ray SED of NGC 1275.
\par
Thus, we consider a constant injection spectrum $Q = Q(E)$ following a power-law distribution,
\begin{equation}
    Q(E) = L_o \left(\frac{E}{E_{ref}}\right)^{-\alpha}
\end{equation} 
for NGC 315, NGC 4261 and NGC 4486 and a log-parabola distribution,
\begin{equation}
    Q(E) = L_o \left(\frac{E}{E_{ref}}\right)^{-\alpha-\beta ln(E/E_{ref})}
\end{equation} 
for NGC 1275, where $E_{ref} = 1$ TeV is the reference energy. The injection spectral index ($\alpha$), the curvature index ($\beta$) and the normalization constant of the spectrum ($L_o$) are free parameters and are determined from the modeling.\\
We calculate the particle spectrum $N = N(E,t)$ at a time t at which the spectrum is assumed to attain a steady state under the continuous injection of particles described by $Q(E)$ and energy losses given by the energy loss rate 
$b = b(E,t)$ using publicly available time-dependent code \texttt{GAMERA\footnote{\url{http://libgamera.github.io/GAMERA/docs/main_page.html}}}(\citealt{hahn}). The code solves 1D transport equation,
\begin{equation}
    \frac{\partial N}{\partial t} = Q(E,t) - \frac{\partial (bN)}{\partial E} - \frac{N}{t_{esc}}
\end{equation}
\\ where $t_{esc}$ is the timescale over which the leptons escape from the emission region.\\
We consider escape time as $t_{esc} = \eta_{esc}\frac{R}{c}$ and $\eta_{esc}$ is considered as a free parameter ($\ge$1).  \\
The code, subsequently, calculates the synchrotron and inverse-Compton emission, which is Doppler boosted by a factor of $\delta^4$ in the observer's frame due to relativistic beaming. $\delta = [\Gamma_b(1-\beta cos\theta)]^{-1}$ is the Doppler factor, $\Gamma_b$ is the bulk Lorentz factor, $\beta$ is the intrinsic speed of the emitting plasma and $\theta$ is the viewing angle of the jet with respect to the line of sight of the observer.

The simulated spectral energy distributions are fitted to the data points by adjusting the parameters given in Table \ref{pc_par}.
 \begin{table*}
\caption{Parameter Values for the best-fit one-zone leptonic SSC model}
\label{pc_par}
\begin{threeparttable}
 \centering
 \begin{tabular}{llllll}
 \hline
 Parameter & Symbol  & NGC 315 & NGC 4261  & NGC 1275 & M87 \\
 \hline
 \hline
 Injection Spectrum Type & $Q_i$ & Power-Law  & Power-Law & Log-Parabola & Power-Law\\ 
 Minimum Electron Lorentz factor & $\gamma_{min}$  & 35 & 190 & 72 & 40\\ 
 Maximum Lorentz factor & $\gamma_{max}$  & 2.5$\times 10^{4}$ & 1.6 $\times 10^{4}$& 5.4$\times 10^{5}$ & 2.8$\times 10^6$\\ 
 Escape time coefficient & $\eta_{esc}$ & 1 & 1 & 1 & 6\\
 Alpha & $\alpha$ & 2.2 & 2.06 & 2.25 & 2.27 \\
 Beta (Curvature Index) & $\beta$ & - & - & 0.015 & -\\
 Lorentz factor & $\Gamma_b$ & 1.5 & 1.5 & 1.8 & 3.3 \\ 
 Doppler Factor & $\delta$ & 1.6\tnote{a} & 1\tnote{a} & 2.3\tnote{b} & 2.3\tnote{c} \\
 Blob Radius (cm) & \emph{R} & 1.1$\times 10^{16}$  &  1$\times 10^{16}$ & 3.13 $\times 10^{17}$& 4.1$\times 10^{15}$  \\
 Magnetic Field (G) & \emph{B}  & 0.21 & 0.21 & 0.07 & 0.24 \\
 Jet power in electrons (ergs/s) & $P_e$& 3.9 $\times 10^{37}$& 4.8 $\times 10^{37}$ &  8.1 $\times 10^{36}$& 1.04 $\times 10^{37}$\\
 Jet power in magnetic field (ergs/s) &$P_B$ & 4.5$\times 10^{40}$ & 3.72 $\times 10^{40}$ & 5.8 $\times 10^{42}$ & 3.9 $\times 10^{40}$\\
 Jet power in cold protons\tnote{d} (ergs/s) & $P_p$ & 4.7$\times 10^{38}$ & 1.12 $\times 10^{38}$ &  2.69 $\times 10^{37}$ & 1.06 $\times 10^{38}$\\
 \hline
 Total jet power (ergs/s) & $P$ & 4.5$\times 10^{40}$ & 3.74 $\times 10^{40}$ & 5.8 $\times 10^{42}$ & 3.96 $\times 10^{40}$\\
 Eddington Jet power (ergs/s) & $P_{edd}$ & 9.9$\times 10^{46}$ & 6.1 $\times 10^{46}$&  3.60 $\times 10^{45}$& 8.17 $\times 10^{47}$\\
 \hline
 \end{tabular}
 \begin{tablenotes}
 \item[a] Adopted from \citet{rani_llagn}.
 \item[b] Adopted from \citet{abdo1275}.
 \item[c] It is close to the value ($\delta=2.8$) used by \citet{fraija}.
 \item[d] Assuming the number of protons is equal to the number of radiating electrons in the jet.
 \end{tablenotes}
 \end{threeparttable}
 \end{table*}
 
 \begin{figure*}
    \centering
    \includegraphics[width=\textwidth]{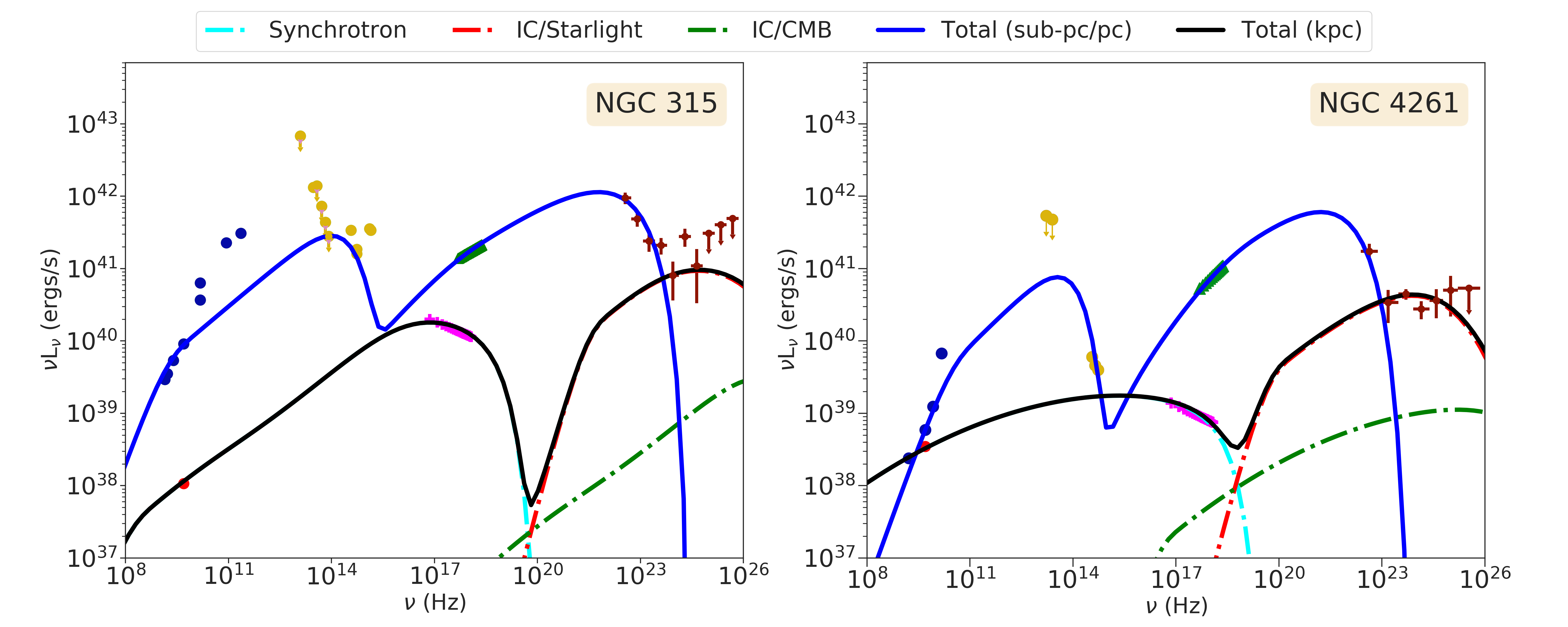}
    \caption{Leptonic model for emission from sub-parsec (blue) and extended jets (black) for NGC 315 and NGC 4261. The radio and X-ray data for the extended jets are shown in red and pink respectively.}
    \label{fig:sedkpc}
\end{figure*}

\begin{table}
\caption{Parameter values for the best-fit one-zone leptonic model for kilo-parsec jet}
\label{tab:kpc_par}
\setlength{\tabcolsep}{0.25\tabcolsep}
\centering
\begin{tabular}{llll}
\hline
 Parameter& Symbol & NGC 315 & NGC 4261 \\
 \hline
 \hline
Starlight energy density & U$_{star}$(ergs/s) & 2.25 $\times 10^{-9}$ & 2.09 $\times 10^{-10}$\\
Dust energy density & U$_{dust}$(ergs/s) & 2.25 $\times 10^{-11}$ & 2.09 $\times 10^{-12}$ \\
CMB energy density & U$_{cmb}$(ergs/s) & 9.$\times 10^{-13}$ & 9.$\times 10^{-13}$ \\
Minimum Lorentz factor & $\gamma_{min}$ & 1400 & 220  \\ 
Maximum Lorentz factor & $\gamma_{max}$& 5.0$\times 10^{8}$ & 5.2$\times 10^{8}$ \\ 
 Spectral index & $\alpha$ & 2.1 & 2.16 \\ 
 Curvature index & $\beta$& 0.025 & 0.084 \\ 
Magnetic field & B (G)& 10$\times 10^{-6}$\  & 5.7$\times 10^{-6}$\\ 
 Radius & R (cm) & 5$\times 10^{21}$ & 1$\times 10^{21}$ \\ 
Escape time factor & $\eta_{esc}$ & 1 & 1\\
 Bulk Lorentz factor & $\Gamma_b$& 1.5  & 1.5 \\ 
 Doppler factor  & $\delta$ & 1.6 & 1\\ 
 Total kpc jet power & P$_{tot}$ (ergs/s)&  2.1 $\times 10^{43}$ & 2.7 $\times 10^{41}$ \\ 
 \hline
\end{tabular}
\end{table}

The total required jet power is calculated as, \begin{equation}
    P_{tot} = \pi R^2 \Gamma_{b}^2 c(U\ensuremath{'}_{e} + U\ensuremath{'}_B + U\ensuremath{'}_p)
\end{equation} where $U\ensuremath{'}_{e}, U\ensuremath{'}_{B}$ and $U\ensuremath{'}_{p}$ are the energy densities of electrons, magnetic field and protons in the comoving frame of jet respectively. These are defined as follows:
\begin{equation}
    U\ensuremath{'}_e  = \frac{1}{V} \times \int_{E_{min}}^{E_{max}} Q(E)E dE
\end{equation}
\begin{equation}
    U\ensuremath{'}_B = \frac{B^2}{8\pi}
\end{equation}
and \begin{equation}
    U\ensuremath{'}_p =n_p\, m_p\,c^2 
\end{equation}
where V is the volume of the emission region, $m_p$ is the mass of proton and $n_p$ is the number density of protons which is equal to the number density of electrons, assuming jet contains equal numbers of electrons and protons to maintain charge neutrality.\\
\subsection{SSC Model for Jet Emission}
As for blazars, the spectral energy distribution of most radio galaxies are well interpreted by a single zone SSC model (for eg. \citealt{abdo_cen}, \citealt{abdo09}). Within this framework, radio to optical photons are  produced by synchrotron radiation of non-thermal electron population in the magnetic field and X-ray to higher energy photons are produced by up scattering of the synchrotron photons by the same electron population. While this model can explain the broadband SED of NGC 1275 and NGC 4486 up to 0.3 TeV and 8 GeV respectively, it fails to do so for NGC 315 beyond 1.6 GeV and NGC 4261 beyond 0.6 GeV. The modeling results can be seen in Figure \ref{fig:sedf}. 
\subsection{Multi-wavelength Emission from Extended jet}
The low angular resolution of instruments at $\gamma-$ray energies does not allow to distinguish between jet and the extended jet emission. We invoke the emission from the kilo-parsec scale jets of NGC 315 and NGC 4261, since the SSC emission from their sub-parsec scale jets cannot fit the $\gamma-$ray data points beyond 1.6 GeV and 0.6 GeV energy respectively. \\
The radio and X-ray photon flux from the extended jets of NGC 315 and NGC 4261 can be well fitted with synchrotron emission by the relativistic electrons (\citealt{worrall315}, \citealt{worrall4261}). This implies that the extended jets could be also sources of high and very high energy photons due to inverse Compton scattering of starlight photons from the host galaxy by the relativistic electrons in the extended jets (\citealt{stawarz03}). The starlight energy density at extended jet of NGC 4261 has been adopted from \citet{worrall4261}. For NGC 315, typical energy density of starlight  in kilo-parsec scale jet for FRI radio galaxies has been adopted (\citealt{stawarz06}).  We have also considered IC/CMB emission, as has been suggested for other large scale X-ray jets (\citealt{zacharias}), where CMB photons are up scattered by the relativistic electrons, but it is found to be sub-dominant as compared to IC/starlight emission. 
\par
We calculate the synchrotron, IC/starlight and IC/CMB emission from kilo-parsec scale jets by considering constant escape time of electrons from the emission region as $R/c$ . We assume the same bulk Lorentz factor and viewing angle for the spherical emission region or blob in the kilo-parsec scale jet as that in the sub-parsec scale jet. The results are shown in Figure \ref{fig:sedkpc} and the corresponding parameters are given in Table \ref{tab:kpc_par}.

\section{Summary and Conclusions}
\label{sec:5}
LLAGNs are important to study as their number is expected to be much higher than the high luminosity AGNs and they have been speculated to be acceleration sites of cosmic rays (\citealt{rodrigues}, \citealt{das}). Detection of these sources by gamma-ray detectors is necessary to support this speculation. 
The four LLAGNs NGC 315, NGC 4261, NGC 1275 and NGC 4486 have been detected in $\gamma$-rays before, hence we have selected these sources for a more extended analysis. 
We have analysed the \textit{XMM-Newton} data from NGC 315 and NGC 4486 and \textit{Swift} data from NGC 315. We have also analysed the \textit{Fermi-LAT} data for all the 
 four LLAGNs NGC 315, NGC 4261, NGC 1275 and NGC 4486 for a period of 12 years (2008-2020). We have combined the archival multi-wavelength data of these sources with our analysed data to build the broadband spectral energy distributions of these sources (see Figure \ref{fig:315}, \ref{fig:4261}, \ref{fig:1275} and \ref{fig:m87}). We have found the best fitted models for the \textit{Fermi-LAT} data (see Figure \ref{fig:fer}) and the parameters of the fitted models are given in Table \ref{fermi_sed}. 
  \par
   For NGC 4486, we find the best-fit describing the  \fermi~ SED is a power-law which is consistent with the best-fit obtained for the combined \fermi~ and \textit{MAGIC} data (\citealt{magic}) for 2012-2015, hence we have used a power-law electron distribution in our modeling to fit the multi-wavelength SED.
   \par
   The gamma-ray data points of NGC 315 and NGC 4261 are also found to be well represented by power-law spectral model, while for NGC 1275 log-parabola distribution gives a better fit to the gamma-ray data. We have considered SSC emission from sub-parsec scale jets of these sources to fit the multi-wavelength data points (see Figure \ref{fig:sedf}) and the corresponding values of the parameters of our model are given in Table \ref{pc_par}. We have included the emission from the extended kilo-parsec scale jets of NGC 315 and NGC 4261 to explain the gamma-ray data points at higher energies (see Figure \ref{fig:sedkpc}), the corresponding values of the model parameters are given in Table \ref{tab:kpc_par}.
 \par
 \citet{rani_llagn} analysed 10.25 years of \fermi~ data and simulated the multi-wavelength SED of NGC 315 and NGC 4261 using one zone SSC model to compare with the observational data. The radio flux estimated in their model is lower than the observed flux for both NGC 315 and NGC 4261. 
 \par
 \citet{abdo1275} used one zone SSC model to fit the multi-wavelength data from NGC 1275, however their model 
 gives higher X-ray flux compared to the observed flux.
 \par 
For NGC 4486, a multizone model has also been proposed to explain the radio to X-ray emission, which gives a lower $\gamma$-ray flux (\citealt{lucchini}) compared to the observed flux.
\par
 While ADAF or jet dominance for X-ray emission from LLAGNs remains under debate, we show that synchrotron and SSC emission from relativistic electrons in sub-parsec scale jets can explain the observed multi-wavelength data from NGC 1275, up to 1.6 GeV from NGC 315, up to 0.6 GeV from NGC 4261 and up to 8 GeV from NGC 4486. At higher energy inverse Compton scattering of starlight photons by electrons accelerated in kilo-parsec scale jets of NGC 315 and NGC 4261 can explain the observed gamma-ray flux. The maximum value of Lorentz factor of the order of 10$^{8}$ obtained to fit the X-ray and gamma-ray data, indicates that the electrons are accelerated to ultra-relativistic energies in the kilo-parsec jets. Such high Lorentz factors also favour the synchrotron origin of X-rays from these jets as suggested before by \citet{worrall315}, \citet{worrall4261}. It is also noted that strong emission of starlight photons from the host galaxy is required for these objects to be seen in GeV $\gamma-$ rays. Kilo-parsec scale jet is also present in NGC 4486, whose radio to X-ray emission has been modeled earlier with synchrotron emission of relativistic electrons (\citealt{2018sun}). 
 These relativistic electrons may also emit by inverse Compton mechanism and contribute to the observed gamma ray flux. Hadronic model has been used earlier to explain the high energy gamma ray data (\citealt{2014marinelli}, \citealt{fraija}). A detailed modeling of NGC 4486 to explain the high energy gamma ray data is beyond the scope of this paper.
 \par
 
 Centaurus A is also considered to be a LLAGN as the luminosity of its H${\alpha}$ emission is less than $10^{40}$ ergs/s (\citealt{2011cena}).
 Earlier, High Energy Stereoscopic System (H.E.S.S.) reported extended TeV gamma ray emission from this LLAGN (\citealt{2020Nature}). They explained the very high energy gamma ray emission as inverse Compton scattering of mainly the dust photons by ultra-relativistic electrons in kilo-parsec scale jet of this source. The extended X-ray emission from kilo-parsec scale jet of this source is explained as synchrotron emission of ultra-relativistic electrons. 
 The kilo-parsec scale jets may also be the acceleration sites of protons and heavy nuclei, which may contribute to the observed spectrum of ultrahigh energy cosmic rays. More observations of high energy gamma rays from nearby LLAGNs would be useful to understand their role as cosmic particle accelerators. 

\section*{Acknowledgement}
The authors would like to thank the referee for the valuable comments and suggestions to improve the manuscript. This research has made use of archival data (from \swift~ and \xmm~ telescopes) and software/tools provided by NASA's High Energy Astrophysics Science Archive Research Center (HEASARC)\footnote{\url{https://heasarc.gsfc.nasa.gov/}}, which is a service of the Astrophysics Science Division at NASA/GSFC. This work has also made use of public \fermi~ data obtained from Fermi Science Support Center (FSSC), provided by NASA Goddard Space Flight Center. R.P. acknowledges the support of the Polish Funding Agency National Science Centre, project 2017/26/A/ST9/00756 (MAESTRO 9), and MNiSW grant DIR/WK/2018/12. 
\software{HEAsoft (v6.26.1; \url{https://heasarc.gsfc.nasa.gov/docs/software/heasoft/}), SAS (v18.0; \citealt{sas}), XSPEC (v12.10.0f; \citealt{xspec}) , fermipy (v1.0.0; \citealt{wood_fermipy}), GAMERA (\citealt{hahn})}

\bibliography{llagn}{}

\begin{thebibliography}{}
\expandafter\ifx\csname natexlab\endcsname\relax\def\natexlab#1{#1}\fi
\providecommand{\url}[1]{\href{#1}{#1}}
\providecommand{\dodoi}[1]{doi:~\href{http://doi.org/#1}{\nolinkurl{#1}}}
\providecommand{\doeprint}[1]{\href{http://ascl.net/#1}{\nolinkurl{http://ascl.net/#1}}}
\providecommand{\doarXiv}[1]{\href{https://arxiv.org/abs/#1}{\nolinkurl{https://arxiv.org/abs/#1}}}

\bibitem[{{Abdo} {et~al.}(2009{\natexlab{a}}){Abdo}, {Ackermann}, {Ajello},
  {Asano}, {Baldini}, {Ballet}, {Barbiellini}, {Bastieri}, {Baughman},
  {Bechtol}, {Bellazzini}, {Blandford}, {Bloom}, {Bonamente}, {Borgland},
  {Bregeon}, {Brez}, {Brigida}, {Bruel}, {Burnett}, {Caliandro}, {Cameron},
  {Caraveo}, {Casandjian}, {Cavazzuti}, {Cecchi}, {Celotti}, {Chekhtman},
  {Cheung}, {Chiang}, {Ciprini}, {Claus}, {Cohen-Tanugi}, {Colafrancesco},
  {Cominsky}, {Conrad}, {Costamante}, {Dermer}, {de Angelis}, {de Palma},
  {Digel}, {Donato}, {do Couto e Silva}, {Drell}, {Dubois}, {Dumora},
  {Farnier}, {Favuzzi}, {Finke}, {Focke}, {Frailis}, {Fukazawa}, {Funk},
  {Fusco}, {Gargano}, {Georganopoulos}, {Germani}, {Giebels}, {Giglietto},
  {Giordano}, {Glanzman}, {Grenier}, {Grondin}, {Grove}, {Guillemot},
  {Guiriec}, {Hanabata}, {Harding}, {Hartman}, {Hayashida}, {Hays}, {Hughes},
  {J{\'o}hannesson}, {Johnson}, {Johnson}, {Johnson}, {Kadler}, {Kamae},
  {Kanai}, {Katagiri}, {Kataoka}, {Kawai}, {Kerr}, {Kn{\"o}dlseder}, {Kuehn},
  {Kuss}, {Latronico}, {Lemoine-Goumard}, {Longo}, {Loparco}, {Lott},
  {Lovellette}, {Lubrano}, {Madejski}, {Makeev}, {Mazziotta}, {McEnery},
  {Meurer}, {Michelson}, {Mitthumsiri}, {Mizuno}, {Moiseev}, {Monte},
  {Monzani}, {Morselli}, {Moskalenko}, {Murgia}, {Nakamori}, {Nolan}, {Norris},
  {Nuss}, {Ohsugi}, {Omodei}, {Orlando}, {Ormes}, {Paneque}, {Panetta},
  {Parent}, {Pepe}, {Pesce-Rollins}, {Piron}, {Porter}, {Rain{\`o}}, {Razzano},
  {Reimer}, {Reimer}, {Reposeur}, {Ritz}, {Rodriguez}, {Romani}, {Ryde},
  {Sadrozinski}, {Sambruna}, {Sanchez}, {Sander}, {Sato}, {Parkinson},
  {Sgr{\`o}}, {Smith}, {Smith}, {Spandre}, {Spinelli}, {Starck}, {Strickman},
  {Strong}, {Suson}, {Tajima}, {Takahashi}, {Takahashi}, {Tanaka}, {Taylor},
  {Thayer}, {Thompson}, {Torres}, {Tosti}, {Uchiyama}, {Usher}, {Vilchez},
  {Vitale}, {Waite}, {Wood}, {Ylinen}, {Ziegler}, {Aller}, {Aller},
  {Kellermann}, {Kovalev}, {Kovalev}, {Lister}, \& {Pushkarev}}]{abdo1275}
{Abdo}, A.~A., {Ackermann}, M., {Ajello}, M., {et~al.} 2009{\natexlab{a}},
  \apj, 699, 31, \dodoi{10.1088/0004-637X/699/1/31}

\bibitem[{{Abdo} {et~al.}(2009{\natexlab{b}}){Abdo}, {Ackermann}, {Ajello},
  {Atwood}, {Axelsson}, {Baldini}, {Ballet}, {Barbiellini}, {Bastieri},
  {Bechtol}, {Bellazzini}, {Berenji}, {Blandford}, {Bloom}, {Bonamente},
  {Borgland}, {Bregeon}, {Brez}, {Brigida}, {Bruel}, {Burnett}, {Caliandro},
  {Cameron}, {Cannon}, {Caraveo}, {Casandjian}, {Cavazzuti}, {Cecchi},
  {{\c{C}}elik}, {Charles}, {Cheung}, {Chiang}, {Ciprini}, {Claus},
  {Cohen-Tanugi}, {Colafrancesco}, {Conrad}, {Costamante}, {Cutini}, {Davis},
  {Dermer}, {de Angelis}, {de Palma}, {Digel}, {Donato}, {Silva}, {Drell},
  {Dubois}, {Dumora}, {Edmonds}, {Farnier}, {Favuzzi}, {Fegan}, {Finke},
  {Focke}, {Fortin}, {Frailis}, {Fukazawa}, {Funk}, {Fusco}, {Gargano},
  {Gasparrini}, {Gehrels}, {Georganopoulos}, {Germani}, {Giebels}, {Giglietto},
  {Giommi}, {Giordano}, {Giroletti}, {Glanzman}, {Godfrey}, {Grenier},
  {Grondin}, {Grove}, {Guillemot}, {Guiriec}, {Hanabata}, {Harding},
  {Hayashida}, {Hays}, {Horan}, {J{\'o}hannesson}, {Johnson}, {Johnson},
  {Johnson}, {Johnson}, {Kamae}, {Katagiri}, {Kataoka}, {Kawai}, {Kerr},
  {Kn{\"o}dlseder}, {Kocian}, {Kuss}, {Lande}, {Latronico}, {Lemoine-Goumard},
  {Longo}, {Loparco}, {Lott}, {Lovellette}, {Lubrano}, {Madejski}, {Makeev},
  {Mazziotta}, {McConville}, {McEnery}, {Meurer}, {Michelson}, {Mitthumsiri},
  {Mizuno}, {Moiseev}, {Monte}, {Monzani}, {Morselli}, {Moskalenko}, {Murgia},
  {Nolan}, {Norris}, {Nuss}, {Ohsugi}, {Omodei}, {Orlando}, {Ormes}, {Ozaki},
  {Paneque}, {Panetta}, {Parent}, {Pelassa}, {Pepe}, {Pesce-Rollins}, {Piron},
  {Porter}, {Rain{\`o}}, {Rando}, {Razzano}, {Reimer}, {Reimer}, {Reposeur},
  {Ritz}, {Rochester}, {Rodriguez}, {Romani}, {Roth}, {Ryde}, {Sadrozinski},
  {Sambruna}, {Sanchez}, {Sander}, {Saz Parkinson}, {Scargle}, {Sgr{\`o}},
  {Shaw}, {Smith}, {Smith}, {Spandre}, {Spinelli}, {Strickman}, {Suson},
  {Tajima}, {Takahashi}, {Tanaka}, {Taylor}, {Thayer}, {Thompson}, {Tibaldo},
  {Torres}, {Tosti}, {Tramacere}, {Uchiyama}, {Usher}, {Vasileiou}, {Vilchez},
  {Waite}, {Wang}, {Winer}, {Wood}, {Ylinen}, {Ziegler}, {Harris}, {Massaro},
  \& {Stawarz}}]{abdo09}
---. 2009{\natexlab{b}}, \apj, 707, 55, \dodoi{10.1088/0004-637X/707/1/55}

\bibitem[{{Abdo} {et~al.}(2010){Abdo}, {Ackermann}, {Ajello}, {Atwood},
  {Baldini}, {Ballet}, {Barbiellini}, {Bastieri}, {Baughman}, {Bechtol},
  {Bellazzini}, {Berenji}, {Blandford}, {Bloom}, {Bonamente}, {Borgland},
  {Bouvier}, {Brandt}, {Bregeon}, {Brez}, {Brigida}, {Bruel}, {Buehler},
  {Buson}, {Caliandro}, {Cameron}, {Cannon}, {Caraveo}, {Carrigan},
  {Casandjian}, {Cavazzuti}, {Cecchi}, {{\c{C}}elik}, {Charles}, {Chekhtman},
  {Cheung}, {Chiang}, {Ciprini}, {Claus}, {Cohen-Tanugi}, {Colafrancesco},
  {Cominsky}, {Conrad}, {Costamante}, {Davis}, {Dermer}, {de Angelis}, {de
  Palma}, {Silva}, {Drell}, {Dubois}, {Dumora}, {Falcone}, {Farnier},
  {Favuzzi}, {Fegan}, {Finke}, {Focke}, {Fortin}, {Frailis}, {Fukazawa},
  {Funk}, {Fusco}, {Gargano}, {Gasparrini}, {Gehrels}, {Georganopoulos},
  {Germani}, {Giebels}, {Giglietto}, {Giommi}, {Giordano}, {Giroletti},
  {Glanzman}, {Godfrey}, {Grandi}, {Grenier}, {Grondin}, {Grove}, {Guillemot},
  {Guiriec}, {Hadasch}, {Harding}, {Hase}, {Hayashida}, {Hays}, {Horan},
  {Hughes}, {Itoh}, {Jackson}, {J{\'o}hannesson}, {Johnson}, {Johnson},
  {Johnson}, {Kadler}, {Kamae}, {Katagiri}, {Kataoka}, {Kawai}, {Kishishita},
  {Kn{\"o}dlseder}, {Kuss}, {Lande}, {Latronico}, {Lee}, {Lemoine-Goumard},
  {Llena Garde}, {Longo}, {Loparco}, {Lott}, {Lovellette}, {Lubrano}, {Makeev},
  {Mazziotta}, {McConville}, {McEnery}, {Michelson}, {Mitthumsiri}, {Mizuno},
  {Moiseev}, {Monte}, {Monzani}, {Morselli}, {Moskalenko}, {Murgia},
  {M{\"u}ller}, {Nakamori}, {Naumann-Godo}, {Nolan}, {Norris}, {Nuss}, {Ohno},
  {Ohsugi}, {Ojha}, {Okumura}, {Omodei}, {Orlando}, {Ormes}, {Ozaki}, {Pagani},
  {Paneque}, {Panetta}, {Parent}, {Pelassa}, {Pepe}, {Pesce-Rollins}, {Piron},
  {Pl{\"o}tz}, {Porter}, {Rain{\`o}}, {Rando}, {Razzano}, {Razzaque}, {Reimer},
  {Reimer}, {Reposeur}, {Ripken}, {Ritz}, {Rodriguez}, {Roth}, {Ryde},
  {Sadrozinski}, {Sanchez}, {Sander}, {Scargle}, {Sgr{\`o}}, {Siskind},
  {Smith}, {Spandre}, {Spinelli}, {Starck}, {Stawarz}, {Strickman}, {Suson},
  {Tajima}, {Takahashi}, {Takahashi}, {Tanaka}, {Thayer}, {Thayer}, {Thompson},
  {Tibaldo}, {Torres}, {Tosti}, {Tramacere}, {Uchiyama}, {Usher},
  {Vandenbroucke}, {Vasileiou}, {Vilchez}, {Vitale}, {Waite}, {Wang}, {Winer},
  {Wood}, {Yang}, {Ylinen}, \& {Ziegler}}]{abdo_cen}
---. 2010, \apj, 719, 1433, \dodoi{10.1088/0004-637X/719/2/1433}

\bibitem[{{Abdollahi} {et~al.}(2020){Abdollahi}, {Acero}, {Ackermann},
  {Ajello}, {Atwood}, {Axelsson}, {Baldini}, {Ballet}, {Barbiellini},
  {Bastieri}, {Becerra Gonzalez}, {Bellazzini}, {Berretta}, {Bissaldi},
  {Blandford}, {Bloom}, {Bonino}, {Bottacini}, {Brandt}, {Bregeon}, {Bruel},
  {Buehler}, {Burnett}, {Buson}, {Cameron}, {Caputo}, {Caraveo}, {Casandjian},
  {Castro}, {Cavazzuti}, {Charles}, {Chaty}, {Chen}, {Cheung}, {Chiaro},
  {Ciprini}, {Cohen-Tanugi}, {Cominsky}, {Coronado-Bl{\'a}zquez}, {Costantin},
  {Cuoco}, {Cutini}, {D'Ammando}, {DeKlotz}, {de la Torre Luque}, {de Palma},
  {Desai}, {Digel}, {Di Lalla}, {Di Mauro}, {Di Venere}, {Dom{\'\i}nguez},
  {Dumora}, {Fana Dirirsa}, {Fegan}, {Ferrara}, {Franckowiak}, {Fukazawa},
  {Funk}, {Fusco}, {Gargano}, {Gasparrini}, {Giglietto}, {Giommi}, {Giordano},
  {Giroletti}, {Glanzman}, {Green}, {Grenier}, {Griffin}, {Grondin}, {Grove},
  {Guiriec}, {Harding}, {Hayashi}, {Hays}, {Hewitt}, {Horan},
  {J{\'o}hannesson}, {Johnson}, {Kamae}, {Kerr}, {Kocevski}, {Kovac'evic'},
  {Kuss}, {Landriu}, {Larsson}, {Latronico}, {Lemoine-Goumard}, {Li},
  {Liodakis}, {Longo}, {Loparco}, {Lott}, {Lovellette}, {Lubrano}, {Madejski},
  {Maldera}, {Malyshev}, {Manfreda}, {Marchesini}, {Marcotulli},
  {Mart{\'\i}-Devesa}, {Martin}, {Massaro}, {Mazziotta}, {McEnery}, {Mereu},
  {Meyer}, {Michelson}, {Mirabal}, {Mizuno}, {Monzani}, {Morselli},
  {Moskalenko}, {Negro}, {Nuss}, {Ojha}, {Omodei}, {Orienti}, {Orlando},
  {Ormes}, {Palatiello}, {Paliya}, {Paneque}, {Pei}, {Pe{\~n}a-Herazo},
  {Perkins}, {Persic}, {Pesce-Rollins}, {Petrosian}, {Petrov}, {Piron}, {Poon},
  {Porter}, {Principe}, {Rain{\`o}}, {Rando}, {Razzano}, {Razzaque}, {Reimer},
  {Reimer}, {Remy}, {Reposeur}, {Romani}, {Saz Parkinson}, {Schinzel},
  {Serini}, {Sgr{\`o}}, {Siskind}, {Smith}, {Spandre}, {Spinelli}, {Strong},
  {Suson}, {Tajima}, {Takahashi}, {Tak}, {Thayer}, {Thompson}, {Tibaldo},
  {Torres}, {Torresi}, {Valverde}, {Van Klaveren}, {van Zyl}, {Wood},
  {Yassine}, \& {Zaharijas}}]{4fgl-cat}
{Abdollahi}, S., {Acero}, F., {Ackermann}, M., {et~al.} 2020, \apjs, 247, 33,
  \dodoi{10.3847/1538-4365/ab6bcb}

\bibitem[{{Abramowicz} {et~al.}(1988){Abramowicz}, {Czerny}, {Lasota}, \&
  {Szuszkiewicz}}]{abram}
{Abramowicz}, M.~A., {Czerny}, B., {Lasota}, J.~P., \& {Szuszkiewicz}, E. 1988,
  \apj, 332, 646, \dodoi{10.1086/166683}

\bibitem[{{Acciari} {et~al.}(2008){Acciari}, {Beilicke}, {Blaylock},
  {Bradbury}, {Buckley}, {Bugaev}, {Butt}, {Celik}, {Cesarini}, {Ciupik},
  {Cogan}, {Colin}, {Cui}, {Daniel}, {Duke}, {Ergin}, {Falcone}, {Fegan},
  {Finley}, {Finnegan}, {Fortin}, {Fortson}, {Gibbs}, {Gillanders}, {Grube},
  {Guenette}, {Gyuk}, {Hanna}, {Hays}, {Holder}, {Horan}, {Hughes}, {Hui},
  {Humensky}, {Imran}, {Kaaret}, {Kertzman}, {Kieda}, {Kildea}, {Konopelko},
  {Krawczynski}, {Krennrich}, {Lang}, {LeBohec}, {Lee}, {Maier}, {McCann},
  {McCutcheon}, {Millis}, {Moriarty}, {Mukherjee}, {Nagai}, {Ong}, {Pandel},
  {Perkins}, {Pohl}, {Quinn}, {Ragan}, {Reynolds}, {Rose}, {Schroedter},
  {Sembroski}, {Smith}, {Steele}, {Swordy}, {Syson}, {Toner}, {Valcarcel},
  {Vassiliev}, {Wakely}, {Ward}, {Weekes}, {Weinstein}, {White}, {Williams},
  {Wissel}, {Wood}, \& {Zitzer}}]{ac08}
{Acciari}, V.~A., {Beilicke}, M., {Blaylock}, G., {et~al.} 2008, \apj, 679,
  397, \dodoi{10.1086/587458}

\bibitem[{{Acciari} {et~al.}(2009){Acciari}, {Aliu}, {Arlen}, {Bautista},
  {Beilicke}, {Benbow}, {Bradbury}, {Buckley}, {Bugaev}, {Butt}, {Byrum},
  {Cannon}, {Celik}, {Cesarini}, {Chow}, {Ciupik}, {Cogan}, {Cui},
  {Dickherber}, {Fegan}, {Finley}, {Fortin}, {Fortson}, {Furniss}, {Gall},
  {Gillanders}, {Grube}, {Guenette}, {Gyuk}, {Hanna}, {Holder}, {Horan}, {Hui},
  {Humensky}, {Imran}, {Kaaret}, {Karlsson}, {Kieda}, {Kildea}, {Konopelko},
  {Krawczynski}, {Krennrich}, {Lang}, {LeBohec}, {Maier}, {McCann},
  {McCutcheon}, {Millis}, {Moriarty}, {Ong}, {Otte}, {Pandel}, {Perkins},
  {Petry}, {Pohl}, {Quinn}, {Ragan}, {Reyes}, {Reynolds}, {Roache}, {Roache},
  {Rose}, {Schroedter}, {Sembroski}, {Smith}, {Swordy}, {Theiling}, {Toner},
  {Varlotta}, {Vincent}, {Wakely}, {Ward}, {Weekes}, {Weinstein}, {Williams},
  {Wissel}, {Wood}, {Walker}, {Davies}, {Hardee}, {Junor}, {Ly}, {Aharonian},
  {Akhperjanian}, {Anton}, {Barres de Almeida}, {Bazer-Bachi}, {Becherini},
  {Behera}, {Bernl{\"o}hr}, {Bochow}, {Boisson}, {Bolmont}, {Borrel},
  {Brucker}, {Brun}, {Brun}, {B{\"u}hler}, {Bulik}, {B{\"u}sching},
  {Boutelier}, {Chadwick}, {Charbonnier}, {Chaves}, {Cheesebrough}, {Chounet},
  {Clapson}, {Coignet}, {Dalton}, {Daniel}, {Davids}, {Degrange}, {Deil},
  {Dickinson}, {Djannati-Ata{\"\i}}, {Domainko}, {Drury}, {Dubois}, {Dubus},
  {Dyks}, {Dyrda}, {Egberts}, {Emmanoulopoulos}, {Espigat}, {Farnier},
  {Feinstein}, {Fiasson}, {F{\"o}rster}, {Fontaine}, {F{\"u}{\ss}ling},
  {Gabici}, {Gallant}, {G{\'e}rard}, {Gerbig}, {Giebels}, {Glicenstein},
  {Gl{\"u}ck}, {Goret}, {G{\"o}hring}, {Hauser}, {Hauser}, {Heinz},
  {Heinzelmann}, {Henri}, {Hermann}, {Hinton}, {Hoffmann}, {Hofmann},
  {Holleran}, {Hoppe}, {Horns}, {Jacholkowska}, {de Jager}, {Jahn}, {Jung},
  {Katarzy{\'n}ski}, {Katz}, {Kaufmann}, {Kendziorra}, {Kerschhaggl},
  {Khangulyan}, {Kh{\'e}lifi}, {Keogh}, {Klu{\'z}niak}, {Kneiske}, {Komin},
  {Kosack}, {Lamanna}, {Lenain}, {Lohse}, {Marandon}, {Martin},
  {Martineau-Huynh}, {Marcowith}, {Maurin}, {McComb}, {Medina}, {Moderski},
  {Moulin}, {Naumann-Godo}, {de Naurois}, {Nedbal}, {Nekrassov}, {Nicholas},
  {Niemiec}, {Nolan}, {Ohm}, {Olive}, {O{\~n}a de Wilhelmi}, {Orford},
  {Ostrowski}, {Panter}, {Paz Arribas}, {Pedaletti}, {Pelletier}, {Petrucci},
  {Pita}, {P{\"u}hlhofer}, {Punch}, {Quirrenbach}, {Raubenheimer}, {Raue},
  {Rayner}, {Renaud}, {Rieger}, {Ripken}, {Rob}, {Rosier-Lees}, {Rowell},
  {Rudak}, {Rulten}, {Ruppel}, {Sahakian}, {Santangelo}, {Schlickeiser},
  {Sch{\"o}ck}, {Schr{\"o}der}, {Schwanke}, {Schwarzburg}, {Schwemmer},
  {Shalchi}, {Sikora}, {Skilton}, {Sol}, {Spangler}, {Stawarz}, {Steenkamp},
  {Stegmann}, {Stinzing}, {Superina}, {Szostek}, {Tam}, {Tavernet}, {Terrier},
  {Tibolla}, {Tluczykont}, {van Eldik}, {Vasileiadis}, {Venter}, {Venter},
  {Vialle}, {Vincent}, {Vivier}, {V{\"o}lk}, {Volpe}, {Wagner}, {Ward},
  {Zdziarski}, {Zech}, {Anderhub}, {Antonelli}, {Antoranz}, {Backes},
  {Baixeras}, {Balestra}, {Barrio}, {Bastieri}, {Becerra Gonz{\'a}lez},
  {Becker}, {Bednarek}, {Berger}, {Bernardini}, {Biland}, {Bock}, {Bonnoli},
  {Bordas}, {Tridon}, {Bosch-Ramon}, {Bose}, {Braun}, {Bretz}, {Britvitch},
  {Camara}, {Carmona}, {Commichau}, {Contreras}, {Cortina}, {Costado},
  {Covino}, {Curtef}, {Dazzi}, {De Angelis}, {de Cea del Pozo}, {Delgado
  Mendez}, {De los Reyes}, {De Lotto}, {De Maria}, {De Sabata}, {Dominguez},
  {Dorner}, {Doro}, {Elsaesser}, {Errando}, {Ferenc}, {Fern{\'a}ndez}, {Firpo},
  {Fonseca}, {Font}, {Galante}, {Garc{\'\i}a L{\'o}pez}, {Garczarczyk}, {Gaug},
  {Goebel}, {Hadasch}, {Hayashida}, {Herrero}, {Hildebrand},
  {H{\"o}hne-M{\"o}nch}, {Hose}, {Hsu}, {Jogler}, {Kranich}, {La Barbera},
  {Laille}, {Leonardo}, {Lindfors}, {Lombardi}, {Longo}, {L{\'o}pez}, {Lorenz},
  {Majumdar}, {Maneva}, {Mankuzhiyil}, {Mannheim}, {Maraschi}, {Mariotti},
  {Mart{\'\i}nez}, {Mazin}, {Meucci}, {Miranda}, {Mirzoyan}, {Miyamoto},
  {Mold{\'o}n}, {Moles}, {Moralejo}, {Nieto}, {Nilsson}, {Ninkovic}, {Oya},
  {Paoletti}, {Paredes}, {Pasanen}, {Pascoli}, {Pauss}, {Pegna},
  {Perez-Torres}, {Persic}, {Peruzzo}, {Prada}, {Prandini}, {Puchades},
  {Reichardt}, {Rhode}, {Rib{\'o}}, {Rico}, {Rissi}, {Robert}, {R{\"u}gamer},
  {Saggion}, {Saito}, {Salvati}, {Sanchez-Conde}, {Satalecka}, {Scalzotto},
  {Scapin}, {Schweizer}, {Shayduk}, {Shore}, {Sidro}, {Sierpowska-Bartosik},
  {Sillanp{\"a}{\"a}}, {Sitarek}, {Sobczynska}, {Spanier}, {Stamerra}, {Stark},
  {Takalo}, {Tavecchio}, {Temnikov}, {Tescaro}, {Teshima}, {Torres}, {Turini},
  {Vankov}, {Wagner}, {Zabalza}, {Zandanel}, {Zanin}, {Zapatero}, {VERITAS
  Collaboration}, {VLBA 43 GHz M87 Monitoring Team}, {H.~E.~S.~S.
  Collaboration}, \& {MAGIC Collaboration}}]{acciari}
{Acciari}, V.~A., {Aliu}, E., {Arlen}, T., {et~al.} 2009, Science, 325, 444,
  \dodoi{10.1126/science.1175406}

\bibitem[{{Acero} {et~al.}(2015){Acero}, {Ackermann}, {Ajello}, {Albert},
  {Atwood}, {Axelsson}, {Baldini}, {Ballet}, {Barbiellini}, {Bastieri},
  {Belfiore}, {Bellazzini}, {Bissaldi}, {Blandford}, {Bloom}, {Bogart},
  {Bonino}, {Bottacini}, {Bregeon}, {Britto}, {Bruel}, {Buehler}, {Burnett},
  {Buson}, {Caliandro}, {Cameron}, {Caputo}, {Caragiulo}, {Caraveo},
  {Casandjian}, {Cavazzuti}, {Charles}, {Chaves}, {Chekhtman}, {Cheung},
  {Chiang}, {Chiaro}, {Ciprini}, {Claus}, {Cohen-Tanugi}, {Cominsky}, {Conrad},
  {Cutini}, {D'Ammando}, {de Angelis}, {DeKlotz}, {de Palma}, {Desiante},
  {Digel}, {Di Venere}, {Drell}, {Dubois}, {Dumora}, {Favuzzi}, {Fegan},
  {Ferrara}, {Finke}, {Franckowiak}, {Fukazawa}, {Funk}, {Fusco}, {Gargano},
  {Gasparrini}, {Giebels}, {Giglietto}, {Giommi}, {Giordano}, {Giroletti},
  {Glanzman}, {Godfrey}, {Grenier}, {Grondin}, {Grove}, {Guillemot}, {Guiriec},
  {Hadasch}, {Harding}, {Hays}, {Hewitt}, {Hill}, {Horan}, {Iafrate}, {Jogler},
  {J{\'o}hannesson}, {Johnson}, {Johnson}, {Johnson}, {Johnson}, {Kamae},
  {Kataoka}, {Katsuta}, {Kuss}, {La Mura}, {Landriu}, {Larsson}, {Latronico},
  {Lemoine-Goumard}, {Li}, {Li}, {Longo}, {Loparco}, {Lott}, {Lovellette},
  {Lubrano}, {Madejski}, {Massaro}, {Mayer}, {Mazziotta}, {McEnery},
  {Michelson}, {Mirabal}, {Mizuno}, {Moiseev}, {Mongelli}, {Monzani},
  {Morselli}, {Moskalenko}, {Murgia}, {Nuss}, {Ohno}, {Ohsugi}, {Omodei},
  {Orienti}, {Orlando}, {Ormes}, {Paneque}, {Panetta}, {Perkins},
  {Pesce-Rollins}, {Piron}, {Pivato}, {Porter}, {Racusin}, {Rando}, {Razzano},
  {Razzaque}, {Reimer}, {Reimer}, {Reposeur}, {Rochester}, {Romani},
  {Salvetti}, {S{\'a}nchez-Conde}, {Saz Parkinson}, {Schulz}, {Siskind},
  {Smith}, {Spada}, {Spandre}, {Spinelli}, {Stephens}, {Strong}, {Suson},
  {Takahashi}, {Takahashi}, {Tanaka}, {Thayer}, {Thayer}, {Thompson},
  {Tibaldo}, {Tibolla}, {Torres}, {Torresi}, {Tosti}, {Troja}, {Van Klaveren},
  {Vianello}, {Winer}, {Wood}, {Wood}, {Zimmer}, \& {Fermi-LAT
  Collaboration}}]{tsc}
{Acero}, F., {Ackermann}, M., {Ajello}, M., {et~al.} 2015, \apjs, 218, 23,
  \dodoi{10.1088/0067-0049/218/2/23}

\bibitem[{{Agudo} {et~al.}(2014){Agudo}, {Thum}, {G{\'o}mez}, \&
  {Wiesemeyer}}]{agudo}
{Agudo}, I., {Thum}, C., {G{\'o}mez}, J.~L., \& {Wiesemeyer}, H. 2014, \aap,
  566, A59, \dodoi{10.1051/0004-6361/201423366}

\bibitem[{{Aharonian} {et~al.}(2003){Aharonian}, {Akhperjanian}, {Beilicke},
  {Bernl{\"o}hr}, {B{\"o}rst}, {Bojahr}, {Bolz}, {Coarasa}, {Contreras},
  {Cortina}, {Denninghoff}, {Fonseca}, {Girma}, {G{\"o}tting}, {Heinzelmann},
  {Hermann}, {Heusler}, {Hofmann}, {Horns}, {Jung}, {Kankanyan}, {Kestel},
  {Kohnle}, {Konopelko}, {Kornmeyer}, {Kranich}, {Lampeitl}, {Lopez}, {Lorenz},
  {Lucarelli}, {Mang}, {Meyer}, {Mirzoyan}, {Moralejo}, {Ona-Wilhelmi},
  {Panter}, {Plyasheshnikov}, {P{\"u}hlhofer}, {de los Reyes}, {Rhode},
  {Ripken}, {Rowell}, {Sahakian}, {Samorski}, {Schilling}, {Siems},
  {Sobzynska}, {Stamm}, {Tluczykont}, {Vitale}, {V{\"o}lk}, {Wiedner}, \&
  {Wittek}}]{ah2003}
{Aharonian}, F., {Akhperjanian}, A., {Beilicke}, M., {et~al.} 2003, \aap, 403,
  L1, \dodoi{10.1051/0004-6361:20030372}

\bibitem[{{Aharonian} {et~al.}(2006){Aharonian}, {Akhperjanian}, {Bazer-Bachi},
  {Beilicke}, {Benbow}, {Berge}, {Bernl{\"o}hr}, {Boisson}, {Bolz}, {Borrel},
  {Braun}, {Brown}, {B{\"u}hler}, {B{\"u}sching}, {Carrigan}, {Chadwick},
  {Chounet}, {Coignet}, {Cornils}, {Costamante}, {Degrange}, {Dickinson},
  {Djannati-Ata{\"\i}}, {Drury}, {Dubus}, {Egberts}, {Emmanoulopoulos},
  {Espigat}, {Feinstein}, {Ferrero}, {Fiasson}, {Fontaine}, {Funk}, {Funk},
  {F{\"u}{\ss}ling}, {Gallant}, {Giebels}, {Glicenstein}, {Goret},
  {Hadjichristidis}, {Hauser}, {Hauser}, {Heinzelmann}, {Henri}, {Hermann},
  {Hinton}, {Hoffmann}, {Hofmann}, {Holleran}, {Hoppe}, {Horns},
  {Jacholkowska}, {de Jager}, {Kendziorra}, {Kerschhaggl}, {Kh{\'e}lifi},
  {Komin}, {Konopelko}, {Kosack}, {Lamanna}, {Latham}, {Le Gallou},
  {Lemi{\`e}re}, {Lemoine-Goumard}, {Lenain}, {Lohse}, {Martin},
  {Martineau-Huynh}, {Marcowith}, {Masterson}, {Maurin}, {McComb}, {Moulin},
  {de Naurois}, {Nedbal}, {Nolan}, {Noutsos}, {Orford}, {Osborne}, {Ouchrif},
  {Panter}, {Pelletier}, {Pita}, {P{\"u}hlhofer}, {Punch}, {Ranchon},
  {Raubenheimer}, {Raue}, {Rayner}, {Reimer}, {Ripken}, {Rob}, {Rolland},
  {Rosier-Lees}, {Rowell}, {Sahakian}, {Santangelo}, {Saug{\'e}}, {Schlenker},
  {Schlickeiser}, {Schr{\"o}der}, {Schwanke}, {Schwarzburg}, {Schwemmer},
  {Shalchi}, {Sol}, {Spangler}, {Spanier}, {Steenkamp}, {Stegmann}, {Superina},
  {Tam}, {Tavernet}, {Terrier}, {Tluczykont}, {van Eldik}, {Vasileiadis},
  {Venter}, {Vialle}, {Vincent}, {V{\"o}lk}, {Wagner}, \& {Ward}}]{ah06}
{Aharonian}, F., {Akhperjanian}, A.~G., {Bazer-Bachi}, A.~R., {et~al.} 2006,
  Science, 314, 1424, \dodoi{10.1126/science.1134408}

\bibitem[{{Ajello} {et~al.}(2009){Ajello}, {Rebusco}, {Cappelluti}, {Reimer},
  {B{\"o}hringer}, {Greiner}, {Gehrels}, {Tueller}, \& {Moretti}}]{ajello}
{Ajello}, M., {Rebusco}, P., {Cappelluti}, N., {et~al.} 2009, \apj, 690, 367,
  \dodoi{10.1088/0004-637X/690/1/367}

\bibitem[{{Ajello} {et~al.}(2020){Ajello}, {Angioni}, {Axelsson}, {Ballet},
  {Barbiellini}, {Bastieri}, {Becerra Gonzalez}, {Bellazzini}, {Bissaldi},
  {Bloom}, {Bonino}, {Bottacini}, {Bruel}, {Buson}, {Cafardo}, {Cameron},
  {Cavazzuti}, {Chen}, {Cheung}, {Ciprini}, {Costantin}, {Cutini}, {D'Ammando},
  {de la Torre Luque}, {de Menezes}, {de Palma}, {Desai}, {Di Lalla}, {Di
  Venere}, {Dom{\'\i}nguez}, {Dirirsa}, {Ferrara}, {Finke}, {Franckowiak},
  {Fukazawa}, {Funk}, {Fusco}, {Gargano}, {Garrappa}, {Gasparrini},
  {Giglietto}, {Giordano}, {Giroletti}, {Green}, {Grenier}, {Guiriec},
  {Harita}, {Hays}, {Horan}, {Itoh}, {J{\'o}hannesson}, {Kovac'evic'},
  {Krauss}, {Kreter}, {Kuss}, {Larsson}, {Leto}, {Li}, {Liodakis}, {Longo},
  {Loparco}, {Lott}, {Lovellette}, {Lubrano}, {Madejski}, {Maldera},
  {Manfreda}, {Mart{\'\i}-Devesa}, {Massaro}, {Mazziotta}, {Mereu}, {Meyer},
  {Migliori}, {Mirabal}, {Mizuno}, {Monzani}, {Morselli}, {Moskalenko},
  {Negro}, {Nemmen}, {Nuss}, {Ojha}, {Ojha}, {Omodei}, {Orienti}, {Orlando},
  {Ormes}, {Paliya}, {Pei}, {Pe{\~n}a-Herazo}, {Persic}, {Pesce-Rollins},
  {Petrov}, {Piron}, {Poon}, {Principe}, {Rain{\`o}}, {Rando}, {Rani},
  {Razzano}, {Razzaque}, {Reimer}, {Reimer}, {Schinzel}, {Serini}, {Sgr{\`o}},
  {Siskind}, {Spandre}, {Spinelli}, {Suson}, {Tachibana}, {Thompson}, {Torres},
  {Torresi}, {Troja}, {Valverde}, {van Zyl}, \& {Yassine}}]{4fgl}
{Ajello}, M., {Angioni}, R., {Axelsson}, M., {et~al.} 2020, \apj, 892, 105,
  \dodoi{10.3847/1538-4357/ab791e}

\bibitem[{{Albert} {et~al.}(2008){Albert}, {Aliu}, {Anderhub}, {Antonelli},
  {Antoranz}, {Backes}, {Baixeras}, {Barrio}, {Bartko}, {Bastieri}, {Becker},
  {Bednarek}, {Berger}, {Bernardini}, {Bigongiari}, {Biland}, {Bock},
  {Bonnoli}, {Bordas}, {Bosch-Ramon}, {Bretz}, {Britvitch}, {Camara},
  {Carmona}, {Chilingarian}, {Commichau}, {Contreras}, {Cortina}, {Costado},
  {Covino}, {Curtef}, {Dazzi}, {De Angelis}, {De Cea del Pozo}, {de los Reyes},
  {De Lotto}, {De Maria}, {De Sabata}, {Delgado Mendez}, {Dominguez}, {Dorner},
  {Doro}, {Errando}, {Fagiolini}, {Ferenc}, {Fern{\'a}ndez}, {Firpo},
  {Fonseca}, {Font}, {Galante}, {Garc{\'\i}a L{\'o}pez}, {Garczarczyk}, {Gaug},
  {Goebel}, {Hayashida}, {Herrero}, {H{\"o}hne}, {Hose}, {Hsu}, {Huber},
  {Jogler}, {Kranich}, {La Barbera}, {Laille}, {Leonardo}, {Lindfors},
  {Lombardi}, {Longo}, {L{\'o}pez}, {Lorenz}, {Majumdar}, {Maneva},
  {Mankuzhiyil}, {Mannheim}, {Maraschi}, {Mariotti}, {Mart{\'\i}nez}, {Mazin},
  {Meucci}, {Meyer}, {Miranda}, {Mirzoyan}, {Mizobuchi}, {Moles}, {Moralejo},
  {Nieto}, {Nilsson}, {Ninkovic}, {Otte}, {Oya}, {Panniello}, {Paoletti},
  {Paredes}, {Pasanen}, {Pascoli}, {Pauss}, {Pegna}, {Perez-Torres}, {Persic},
  {Peruzzo}, {Piccioli}, {Prada}, {Prandini}, {Puchades}, {Raymers}, {Rhode},
  {Rib{\'o}}, {Rico}, {Rissi}, {Robert}, {R{\"u}gamer}, {Saggion}, {Saito},
  {Salvati}, {Sanchez-Conde}, {Sartori}, {Satalecka}, {Scalzotto}, {Scapin},
  {Schweizer}, {Shayduk}, {Shinozaki}, {Shore}, {Sidro}, {Sierpowska-Bartosik},
  {Sillanp{\"a}{\"a}}, {Sobczynska}, {Spanier}, {Stamerra}, {Stark}, {Takalo},
  {Tavecchio}, {Temnikov}, {Tescaro}, {Teshima}, {Tluczykont}, {Torres},
  {Turini}, {Vankov}, {Venturini}, {Vitale}, {Wagner}, {Wittek}, {Zabalza},
  {Zandanel}, {Zanin}, \& {Zapatero}}]{al08}
{Albert}, J., {Aliu}, E., {Anderhub}, H., {et~al.} 2008, \apjl, 685, L23,
  \dodoi{10.1086/592348}

\bibitem[{{Aleksi{\'c}} {et~al.}(2012){Aleksi{\'c}}, {Alvarez}, {Antonelli},
  {Antoranz}, {Asensio}, {Backes}, {Barres de Almeida}, {Barrio}, {Bastieri},
  {Becerra Gonz{\'a}lez}, {Bednarek}, {Berger}, {Bernardini}, {Biland},
  {Blanch}, {Bock}, {Boller}, {Bonnoli}, {Borla Tridon}, {Bretz},
  {Ca{\~n}ellas}, {Carmona}, {Carosi}, {Colin}, {Colombo}, {Contreras},
  {Cortina}, {Cossio}, {Covino}, {da Vela}, {Dazzi}, {de Angelis}, {de Caneva},
  {de Cea Del Pozo}, {de Lotto}, {Delgado Mendez}, {Diago Ortega}, {Doert},
  {Dom{\'\i}nguez}, {Dominis Prester}, {Dorner}, {Doro}, {Eisenacher},
  {Elsaesser}, {Ferenc}, {Fonseca}, {Font}, {Fruck}, {Garc{\'\i}a L{\'o}pez},
  {Garczarczyk}, {Garrido}, {Giavitto}, {Godinovi{\'c}}, {Gozzini}, {Hadasch},
  {H{\"a}fner}, {Herrero}, {Hildebrand}, {H{\"o}hne-M{\"o}nch}, {Hose},
  {Hrupec}, {Huber}, {Jogler}, {Kadenius}, {Kellermann}, {Klepser},
  {Kr{\"a}henb{\"u}hl}, {Krause}, {La Barbera}, {Lelas}, {Leonardo},
  {Lewandowska}, {Lindfors}, {Lombardi}, {L{\'o}pez}, {L{\'o}pez-Coto},
  {L{\'o}pez-Oramas}, {Lorenz}, {Makariev}, {Maneva}, {Mankuzhiyil},
  {Mannheim}, {Maraschi}, {Mariotti}, {Mart{\'\i}nez}, {Mazin}, {Meucci},
  {Miranda}, {Mirzoyan}, {Mold{\'o}n}, {Moralejo}, {Munar-Adrover},
  {Niedzwiecki}, {Nieto}, {Nilsson}, {Nowak}, {Orito}, {Paiano}, {Paneque},
  {Paoletti}, {Pardo}, {Paredes}, {Partini}, {Perez-Torres}, {Persic},
  {Peruzzo}, {Pilia}, {Pochon}, {Prada}, {Prada Moroni}, {Prandini}, {Puerto
  Gimenez}, {Puljak}, {Reichardt}, {Reinthal}, {Rhode}, {Rib{\'o}}, {Rico},
  {R{\"u}gamer}, {Saggion}, {Saito}, {Saito}, {Salvati}, {Satalecka},
  {Scalzotto}, {Scapin}, {Schultz}, {Schweizer}, {Shayduk}, {Shore},
  {Sillanp{\"a}{\"a}}, {Sitarek}, {Snidaric}, {Sobczynska}, {Spanier}, {Spiro},
  {Stamatescu}, {Stamerra}, {Steinke}, {Storz}, {Strah}, {Sun}, {Suri{\'c}},
  {Takalo}, {Takami}, {Tavecchio}, {Temnikov}, {Terzi{\'c}}, {Tescaro},
  {Teshima}, {Tibolla}, {Torres}, {Treves}, {Uellenbeck}, {Vogler}, {Wagner},
  {Weitzel}, {Zabalza}, {Zandanel}, {Zanin}, {Pfrommer}, \& {Pinzke}}]{alek12}
{Aleksi{\'c}}, J., {Alvarez}, E.~A., {Antonelli}, L.~A., {et~al.} 2012, \aap,
  539, L2, \dodoi{10.1051/0004-6361/201118668}

\bibitem[{{Antonucci}(1993)}]{ant}
{Antonucci}, R. 1993, \araa, 31, 473,
  \dodoi{10.1146/annurev.aa.31.090193.002353}

\bibitem[{{Arnaud}(1996)}]{xspec}
{Arnaud}, K.~A. 1996, in Astronomical Society of the Pacific Conference Series,
  Vol. 101, Astronomical Data Analysis Software and Systems V, ed. G.~H.
  {Jacoby} \& J.~{Barnes}, 17

\bibitem[{{Asada} {et~al.}(2006){Asada}, {Kameno}, {Shen}, {Horiuchi},
  {Gabuzda}, \& {Inoue}}]{asada}
{Asada}, K., {Kameno}, S., {Shen}, Z.-Q., {et~al.} 2006, \pasj, 58, 261,
  \dodoi{10.1093/pasj/58.2.261}

\bibitem[{{Asmus} {et~al.}(2014){Asmus}, {H{\"o}nig}, {Gandhi}, {Smette}, \&
  {Duschl}}]{asmus}
{Asmus}, D., {H{\"o}nig}, S.~F., {Gandhi}, P., {Smette}, A., \& {Duschl}, W.~J.
  2014, \mnras, 439, 1648, \dodoi{10.1093/mnras/stu041}

\bibitem[{{Atwood} {et~al.}(2013){Atwood}, {Albert}, {Baldini}, {Tinivella},
  {Bregeon}, {Pesce-Rollins}, {Sgr{\`o}}, {Bruel}, {Charles}, {Drlica-Wagner},
  {Franckowiak}, {Jogler}, {Rochester}, {Usher}, {Wood}, {Cohen-Tanugi}, \&
  {Zimmer}}]{pass8}
{Atwood}, W., {Albert}, A., {Baldini}, L., {et~al.} 2013, arXiv e-prints,
  arXiv:1303.3514.
\newblock \doarXiv{1303.3514}

\bibitem[{{Atwood} {et~al.}(2009{\natexlab{a}}){Atwood}, {Abdo}, {Ackermann},
  {Althouse}, {Anderson}, {Axelsson}, {Baldini}, {Ballet}, {Band},
  {Barbiellini}, \& et~al.}]{lat}
{Atwood}, W.~B., {Abdo}, A.~A., {Ackermann}, M., {et~al.} 2009{\natexlab{a}},
  \apj, 697, 1071, \dodoi{10.1088/0004-637X/697/2/1071}

\bibitem[{{Atwood} {et~al.}(2009{\natexlab{b}}){Atwood}, {Abdo}, {Ackermann},
  {Althouse}, {Anderson}, {Axelsson}, {Baldini}, {Ballet}, {Band},
  {Barbiellini}, {Bartelt}, {Bastieri}, {Baughman}, {Bechtol},
  {B{\'e}d{\'e}r{\`e}de}, {Bellardi}, {Bellazzini}, {Berenji}, {Bignami},
  {Bisello}, {Bissaldi}, {Blandford}, {Bloom}, {Bogart}, {Bonamente},
  {Bonnell}, {Borgland}, {Bouvier}, {Bregeon}, {Brez}, {Brigida}, {Bruel},
  {Burnett}, {Busetto}, {Caliandro}, {Cameron}, {Caraveo}, {Carius}, {Carlson},
  {Casandjian}, {Cavazzuti}, {Ceccanti}, {Cecchi}, {Charles}, {Chekhtman},
  {Cheung}, {Chiang}, {Chipaux}, {Cillis}, {Ciprini}, {Claus}, {Cohen-Tanugi},
  {Condamoor}, {Conrad}, {Corbet}, {Corucci}, {Costamante}, {Cutini}, {Davis},
  {Decotigny}, {DeKlotz}, {Dermer}, {de Angelis}, {Digel}, {do Couto e Silva},
  {Drell}, {Dubois}, {Dumora}, {Edmonds}, {Fabiani}, {Farnier}, {Favuzzi},
  {Flath}, {Fleury}, {Focke}, {Funk}, {Fusco}, {Gargano}, {Gasparrini},
  {Gehrels}, {Gentit}, {Germani}, {Giebels}, {Giglietto}, {Giommi}, {Giordano},
  {Glanzman}, {Godfrey}, {Grenier}, {Grondin}, {Grove}, {Guillemot}, {Guiriec},
  {Haller}, {Harding}, {Hart}, {Hays}, {Healey}, {Hirayama}, {Hjalmarsdotter},
  {Horn}, {Hughes}, {J{\'o}hannesson}, {Johansson}, {Johnson}, {Johnson},
  {Johnson}, {Johnson}, {Kamae}, {Katagiri}, {Kataoka}, {Kavelaars}, {Kawai},
  {Kelly}, {Kerr}, {Klamra}, {Kn{\"o}dlseder}, {Kocian}, {Komin}, {Kuehn},
  {Kuss}, {Landriu}, {Latronico}, {Lee}, {Lee}, {Lemoine-Goumard}, {Lionetto},
  {Longo}, {Loparco}, {Lott}, {Lovellette}, {Lubrano}, {Madejski}, {Makeev},
  {Marangelli}, {Massai}, {Mazziotta}, {McEnery}, {Menon}, {Meurer},
  {Michelson}, {Minuti}, {Mirizzi}, {Mitthumsiri}, {Mizuno}, {Moiseev},
  {Monte}, {Monzani}, {Moretti}, {Morselli}, {Moskalenko}, {Murgia},
  {Nakamori}, {Nishino}, {Nolan}, {Norris}, {Nuss}, {Ohno}, {Ohsugi}, {Omodei},
  {Orlando}, {Ormes}, {Paccagnella}, {Paneque}, {Panetta}, {Parent}, {Pearce},
  {Pepe}, {Perazzo}, {Pesce-Rollins}, {Picozza}, {Pieri}, {Pinchera}, {Piron},
  {Porter}, {Poupard}, {Rain{\`o}}, {Rando}, {Rapposelli}, {Razzano}, {Reimer},
  {Reimer}, {Reposeur}, {Reyes}, {Ritz}, {Rochester}, {Rodriguez}, {Romani},
  {Roth}, {Russell}, {Ryde}, {Sabatini}, {Sadrozinski}, {Sanchez}, {Sander},
  {Sapozhnikov}, {Parkinson}, {Scargle}, {Schalk}, {Scolieri}, {Sgr{\`o}},
  {Share}, {Shaw}, {Shimokawabe}, {Shrader}, {Sierpowska-Bartosik}, {Siskind},
  {Smith}, {Smith}, {Spandre}, {Spinelli}, {Starck}, {Stephens}, {Strickman},
  {Strong}, {Suson}, {Tajima}, {Takahashi}, {Takahashi}, {Tanaka}, {Tenze},
  {Tether}, {Thayer}, {Thayer}, {Thompson}, {Tibaldo}, {Tibolla}, {Torres},
  {Tosti}, {Tramacere}, {Turri}, {Usher}, {Vilchez}, {Vitale}, {Wang},
  {Watters}, {Winer}, {Wood}, {Ylinen}, \& {Ziegler}}]{atwood}
---. 2009{\natexlab{b}}, \apj, 697, 1071, \dodoi{10.1088/0004-637X/697/2/1071}

\bibitem[{{Ballet} {et~al.}(2020){Ballet}, {Burnett}, {Digel}, \&
  {Lott}}]{4fgldr}
{Ballet}, J., {Burnett}, T.~H., {Digel}, S.~W., \& {Lott}, B. 2020, arXiv
  e-prints, arXiv:2005.11208.
\newblock \doarXiv{2005.11208}

\bibitem[{{Biretta} {et~al.}(1999){Biretta}, {Sparks}, \&
  {Macchetto}}]{biretta}
{Biretta}, J.~A., {Sparks}, W.~B., \& {Macchetto}, F. 1999, \apj, 520, 621,
  \dodoi{10.1086/307499}

\bibitem[{{Birkinshaw} \& {Davies}(1985)}]{birkin}
{Birkinshaw}, M., \& {Davies}, R.~L. 1985, \apj, 291, 32,
  \dodoi{10.1086/163038}

\bibitem[{{B{\"o}hringer} {et~al.}(2001){B{\"o}hringer}, {Belsole}, {Kennea},
  {Matsushita}, {Molendi}, {Worrall}, {Mushotzky}, {Ehle}, {Guainazzi},
  {Sakelliou}, {Stewart}, {Vestrand}, \& {Dos Santos}}]{bohringer}
{B{\"o}hringer}, H., {Belsole}, E., {Kennea}, J., {et~al.} 2001, \aap, 365,
  L181, \dodoi{10.1051/0004-6361:20000092}

\bibitem[{{Brodatzki} {et~al.}(2011){Brodatzki}, {Pardy}, {Becker}, \&
  {Schlickeiser}}]{2011cena}
{Brodatzki}, K.~A., {Pardy}, D. J.~S., {Becker}, J.~K., \& {Schlickeiser}, R.
  2011, \apj, 736, 98, \dodoi{10.1088/0004-637X/736/2/98}

\bibitem[{{Burrows} {et~al.}(2005){Burrows}, {Hill}, {Nousek}, {Kennea},
  {Wells}, {Osborne}, {Abbey}, {Beardmore}, {Mukerjee}, {Short}, {Chincarini},
  {Campana}, {Citterio}, {Moretti}, {Pagani}, {Tagliaferri}, {Giommi},
  {Capalbi}, {Tamburelli}, {Angelini}, {Cusumano}, {Br{\"a}uninger}, {Burkert},
  \& {Hartner}}]{burrows}
{Burrows}, D.~N., {Hill}, J.~E., {Nousek}, J.~A., {et~al.} 2005, \ssr, 120,
  165, \dodoi{10.1007/s11214-005-5097-2}

\bibitem[{{Canvin} {et~al.}(2005){Canvin}, {Laing}, {Bridle}, \&
  {Cotton}}]{canvin}
{Canvin}, J.~R., {Laing}, R.~A., {Bridle}, A.~H., \& {Cotton}, W.~D. 2005,
  \mnras, 363, 1223, \dodoi{10.1111/j.1365-2966.2005.09537.x}

\bibitem[{{Capetti} {et~al.}(2005){Capetti}, {Verdoes Kleijn}, \&
  {Chiaberge}}]{cap05}
{Capetti}, A., {Verdoes Kleijn}, G., \& {Chiaberge}, M. 2005, \aap, 439, 935,
  \dodoi{10.1051/0004-6361:20041609}

\bibitem[{{Cappellari} {et~al.}(2011){Cappellari}, {Emsellem}, {Krajnovi{\'c}},
  {McDermid}, {Scott}, {Verdoes Kleijn}, {Young}, {Alatalo}, {Bacon}, {Blitz},
  {Bois}, {Bournaud}, {Bureau}, {Davies}, {Davis}, {de Zeeuw}, {Duc},
  {Khochfar}, {Kuntschner}, {Lablanche}, {Morganti}, {Naab}, {Oosterloo},
  {Sarzi}, {Serra}, \& {Weijmans}}]{cap11}
{Cappellari}, M., {Emsellem}, E., {Krajnovi{\'c}}, D., {et~al.} 2011, \mnras,
  413, 813, \dodoi{10.1111/j.1365-2966.2010.18174.x}

\bibitem[{{Cotton} {et~al.}(1999){Cotton}, {Feretti}, {Giovannini}, {Lara}, \&
  {Venturi}}]{cotton}
{Cotton}, W.~D., {Feretti}, L., {Giovannini}, G., {Lara}, L., \& {Venturi}, T.
  1999, \apj, 519, 108, \dodoi{10.1086/307358}

\bibitem[{{Curtis}(1918)}]{curtis}
{Curtis}, H.~D. 1918, Publications of Lick Observatory, 13, 9

\bibitem[{{Das} {et~al.}(2020){Das}, {Gupta}, \& {Razzaque}}]{das}
{Das}, S., {Gupta}, N., \& {Razzaque}, S. 2020, \apj, 889, 149,
  \dodoi{10.3847/1538-4357/ab6131}

\bibitem[{{de Jong} {et~al.}(2015){de Jong}, {Beckmann}, {Soldi}, {Tramacere},
  \& {Gros}}]{dejong}
{de Jong}, S., {Beckmann}, V., {Soldi}, S., {Tramacere}, A., \& {Gros}, A.
  2015, \mnras, 450, 4333, \dodoi{10.1093/mnras/stv927}

\bibitem[{{de Menezes} {et~al.}(2020){de Menezes}, {Nemmen}, {Finke},
  {Almeida}, \& {Rani}}]{rani_llagn}
{de Menezes}, R., {Nemmen}, R., {Finke}, J.~D., {Almeida}, I., \& {Rani}, B.
  2020, \mnras, 492, 4120, \dodoi{10.1093/mnras/staa083}

\bibitem[{{Donato} {et~al.}(2004){Donato}, {Sambruna}, \& {Gliozzi}}]{donato}
{Donato}, D., {Sambruna}, R.~M., \& {Gliozzi}, M. 2004, \apj, 617, 915,
  \dodoi{10.1086/425575}

\bibitem[{{Event Horizon Telescope Collaboration} {et~al.}(2019){Event Horizon
  Telescope Collaboration}, {Akiyama}, {Alberdi}, {Alef}, {Asada}, {Azulay},
  {Baczko}, {Ball}, {Balokovi{\'c}}, {Barrett}, {Bintley}, {Blackburn},
  {Boland}, {Bouman}, {Bower}, {Bremer}, {Brinkerink}, {Brissenden}, {Britzen},
  {Broderick}, {Broguiere}, {Bronzwaer}, {Byun}, {Carlstrom}, {Chael}, {Chan},
  {Chatterjee}, {Chatterjee}, {Chen}, {Chen}, {Cho}, {Christian}, {Conway},
  {Cordes}, {Crew}, {Cui}, {Davelaar}, {De Laurentis}, {Deane}, {Dempsey},
  {Desvignes}, {Dexter}, {Doeleman}, {Eatough}, {Falcke}, {Fish}, {Fomalont},
  {Fraga-Encinas}, {Friberg}, {Fromm}, {G{\'o}mez}, {Galison}, {Gammie},
  {Garc{\'\i}a}, {Gentaz}, {Georgiev}, {Goddi}, {Gold}, {Gu}, {Gurwell},
  {Hada}, {Hecht}, {Hesper}, {Ho}, {Ho}, {Honma}, {Huang}, {Huang}, {Hughes},
  {Ikeda}, {Inoue}, {Issaoun}, {James}, {Jannuzi}, {Janssen}, {Jeter}, {Jiang},
  {Johnson}, {Jorstad}, {Jung}, {Karami}, {Karuppusamy}, {Kawashima},
  {Keating}, {Kettenis}, {Kim}, {Kim}, {Kim}, {Kino}, {Koay}, {Koch}, {Koyama},
  {Kramer}, {Kramer}, {Krichbaum}, {Kuo}, {Lauer}, {Lee}, {Li}, {Li},
  {Lindqvist}, {Liu}, {Liuzzo}, {Lo}, {Lobanov}, {Loinard}, {Lonsdale}, {Lu},
  {MacDonald}, {Mao}, {Markoff}, {Marrone}, {Marscher}, {Mart{\'\i}-Vidal},
  {Matsushita}, {Matthews}, {Medeiros}, {Menten}, {Mizuno}, {Mizuno}, {Moran},
  {Moriyama}, {Moscibrodzka}, {M{\"u}ller}, {Nagai}, {Nagar}, {Nakamura},
  {Narayan}, {Narayanan}, {Natarajan}, {Neri}, {Ni}, {Noutsos}, {Okino},
  {Olivares}, {Oyama}, {{\"O}zel}, {Palumbo}, {Patel}, {Pen}, {Pesce},
  {Pi{\'e}tu}, {Plambeck}, {PopStefanija}, {Porth}, {Prather},
  {Preciado-L{\'o}pez}, {Psaltis}, {Pu}, {Ramakrishnan}, {Rao}, {Rawlings},
  {Raymond}, {Rezzolla}, {Ripperda}, {Roelofs}, {Rogers}, {Ros}, {Rose},
  {Roshanineshat}, {Rottmann}, {Roy}, {Ruszczyk}, {Ryan}, {Rygl},
  {S{\'a}nchez}, {S{\'a}nchez-Arguelles}, {Sasada}, {Savolainen}, {Schloerb},
  {Schuster}, {Shao}, {Shen}, {Small}, {Sohn}, {SooHoo}, {Tazaki}, {Tiede},
  {Tilanus}, {Titus}, {Toma}, {Torne}, {Trent}, {Trippe}, {Tsuda}, {van
  Bemmel}, {van Langevelde}, {van Rossum}, {Wagner}, {Wardle}, {Weintroub},
  {Wex}, {Wharton}, {Wielgus}, {Wong}, {Wu}, {Young}, {Young}, {Younsi},
  {Yuan}, {Yuan}, {Zensus}, {Zhao}, {Zhao}, {Zhu}, {Farah}, {Meyer-Zhao},
  {Michalik}, {Nadolski}, {Nishioka}, {Pradel}, {Primiani}, {Souccar},
  {Vertatschitsch}, \& {Yamaguchi}}]{eht}
{Event Horizon Telescope Collaboration}, {Akiyama}, K., {Alberdi}, A., {et~al.}
  2019, \apjl, 875, L6, \dodoi{10.3847/2041-8213/ab1141}

\bibitem[{{Falcke} {et~al.}(2000){Falcke}, {Nagar}, {Wilson}, \&
  {Ulvestad}}]{falc}
{Falcke}, H., {Nagar}, N.~M., {Wilson}, A.~S., \& {Ulvestad}, J.~S. 2000, \apj,
  542, 197, \dodoi{10.1086/309543}

\bibitem[{{Ferrarese} \& {Ford}(2005)}]{ferrarese}
{Ferrarese}, L., \& {Ford}, H. 2005, \ssr, 116, 523,
  \dodoi{10.1007/s11214-005-3947-6}

\bibitem[{{Ferrarese} {et~al.}(1996){Ferrarese}, {Ford}, \& {Jaffe}}]{ferrar}
{Ferrarese}, L., {Ford}, H.~C., \& {Jaffe}, W. 1996, \apj, 470, 444,
  \dodoi{10.1086/177876}

\bibitem[{{Fraija} \& {Marinelli}(2016)}]{fraija}
{Fraija}, N., \& {Marinelli}, A. 2016, \apj, 830, 81,
  \dodoi{10.3847/0004-637X/830/2/81}

\bibitem[{{Fujita} \& {Nagai}(2017)}]{fujita}
{Fujita}, Y., \& {Nagai}, H. 2017, \mnras, 465, L94,
  \dodoi{10.1093/mnrasl/slw217}

\bibitem[{{Fukazawa} {et~al.}(2015){Fukazawa}, {Finke}, {Stawarz}, {Tanaka},
  {Itoh}, \& {Tokuda}}]{fuka18}
{Fukazawa}, Y., {Finke}, J., {Stawarz}, {\L}., {et~al.} 2015, \apj, 798, 74,
  \dodoi{10.1088/0004-637X/798/2/74}

\bibitem[{{Gabriel} {et~al.}(2004){Gabriel}, {Denby}, {Fyfe}, {Hoar}, {Ibarra},
  {Ojero}, {Osborne}, {Saxton}, {Lammers}, \& {Vacanti}}]{sas}
{Gabriel}, C., {Denby}, M., {Fyfe}, D.~J., {et~al.} 2004, in Astronomical
  Society of the Pacific Conference Series, Vol. 314, Astronomical Data
  Analysis Software and Systems (ADASS) XIII, ed. F.~{Ochsenbein}, M.~G.
  {Allen}, \& D.~{Egret}, 759

\bibitem[{{Ghosal} {et~al.}(2020){Ghosal}, {Tolamatti}, {Singh}, {Yadav},
  {Rannot}, {Tickoo}, {Chandra}, {Goyal}, {Kumar}, {Marandi}, {Agarwal},
  {Kothari}, {Gaur}, {Chouhan}, {Borwankar}, {Dhar}, {Koul}, {Venugopal},
  {Chanchalani}, {Bhat}, {Godiyal}, {Mankuzhiyil}, {Godambe}, {Sahayanathan},
  \& {Sharma}}]{gho20}
{Ghosal}, B., {Tolamatti}, A., {Singh}, K.~K., {et~al.} 2020, \na, 80, 101402,
  \dodoi{10.1016/j.newast.2020.101402}

\bibitem[{{Gliozzi} {et~al.}(2003){Gliozzi}, {Sambruna}, \& {Brandt}}]{gliozzi}
{Gliozzi}, M., {Sambruna}, R.~M., \& {Brandt}, W.~N. 2003, \aap, 408, 949,
  \dodoi{10.1051/0004-6361:20031050}

\bibitem[{{Gonz{\'a}lez-Mart{\'\i}n} {et~al.}(2006){Gonz{\'a}lez-Mart{\'\i}n},
  {Masegosa}, {M{\'a}rquez}, {Guerrero}, \& {Dultzin-Hacyan}}]{gonz06}
{Gonz{\'a}lez-Mart{\'\i}n}, O., {Masegosa}, J., {M{\'a}rquez}, I., {Guerrero},
  M.~A., \& {Dultzin-Hacyan}, D. 2006, \aap, 460, 45,
  \dodoi{10.1051/0004-6361:20054756}

\bibitem[{{Gu} {et~al.}(2007){Gu}, {Huang}, {Wilson}, \& {Fazio}}]{gu07}
{Gu}, Q.~S., {Huang}, J.~S., {Wilson}, G., \& {Fazio}, G.~G. 2007, \apjl, 671,
  L105, \dodoi{10.1086/525018}

\bibitem[{{H.~E.~S.~S. Collaboration} {et~al.}(2020){H.~E.~S.~S.
  Collaboration}, {Abdalla}, {Adam}, {Aharonian}, {Ait Benkhali},
  {Ang{\"u}ner}, {Arakawa}, {Arcaro}, {Armand}, {Ashkar}, {Backes}, {Barbosa
  Martins}, {Barnard}, {Becherini}, {Berge}, {Bernl{\"o}hr}, {Blackwell},
  {B{\"o}ttcher}, {Boisson}, {Bolmont}, {Bonnefoy}, {Bregeon}, {Breuhaus},
  {Brun}, {Brun}, {Bryan}, {B{\"u}chele}, {Bulik}, {Bylund}, {Capasso},
  {Caroff}, {Carosi}, {Casanova}, {Cerruti}, {Chand}, {Chandra}, {Chen},
  {Colafrancesco}, {Cury{\l}o}, {Davids}, {Deil}, {Devin}, {deWilt}, {Dirson},
  {Djannati-Ata{\"\i}}, {Dmytriiev}, {Donath}, {Doroshenko}, {Drury}, {Dyks},
  {Egberts}, {Emery}, {Ernenwein}, {Eschbach}, {Feijen}, {Fegan}, {Fiasson},
  {Fontaine}, {Funk}, {F{\"u}{\ss}ling}, {Gabici}, {Gallant}, {Gat{\'e}},
  {Giavitto}, {Glawion}, {Glicenstein}, {Gottschall}, {Grondin}, {Hahn},
  {Haupt}, {Heinzelmann}, {Henri}, {Hermann}, {Hinton}, {Hofmann}, {Hoischen},
  {Holch}, {Holler}, {Horns}, {Huber}, {Iwasaki}, {Jamrozy}, {Jankowsky},
  {Jankowsky}, {Jardin-Blicq}, {Jung-Richardt}, {Kastendieck},
  {Katarzy{\'n}ski}, {Katsuragawa}, {Katz}, {Khangulyan}, {Kh{\'e}lifi},
  {King}, {Klepser}, {Klu{\'z}niak}, {Komin}, {Kosack}, {Kostunin}, {Kraus},
  {Lamanna}, {Lau}, {Lemi{\`e}re}, {Lemoine-Goumard}, {Lenain}, {Leser},
  {Levy}, {Lohse}, {Lypova}, {Mackey}, {Majumdar}, {Malyshev}, {Marandon},
  {Marcowith}, {Mares}, {Mariaud}, {Mart{\'\i}-Devesa}, {Marx}, {Maurin},
  {Meintjes}, {Mitchell}, {Moderski}, {Mohamed}, {Mohrmann}, {Moore}, {Moulin},
  {Muller}, {Murach}, {Nakashima}, {de Naurois}, {Ndiyavala}, {Niederwanger},
  {Niemiec}, {Oakes}, {O'Brien}, {Odaka}, {Ohm}, {de Ona Wilhelmi},
  {Ostrowski}, {Oya}, {Panter}, {Parsons}, {Perennes}, {Petrucci}, {Peyaud},
  {Piel}, {Pita}, {Poireau}, {Priyana Noel}, {Prokhorov}, {Prokoph},
  {P{\"u}hlhofer}, {Punch}, {Quirrenbach}, {Raab}, {Rauth}, {Reimer}, {Reimer},
  {Remy}, {Renaud}, {Rieger}, {Rinchiuso}, {Romoli}, {Rowell}, {Rudak},
  {Ruiz-Velasco}, {Sahakian}, {Saito}, {Sanchez}, {Santangelo}, {Sasaki},
  {Schlickeiser}, {Sch{\"u}ssler}, {Schulz}, {Schutte}, {Schwanke},
  {Schwemmer}, {Seglar-Arroyo}, {Senniappan}, {Seyffert}, {Shafi},
  {Shiningayamwe}, {Simoni}, {Sinha}, {Sol}, {Specovius}, {Spir-Jacob},
  {Stawarz}, {Steenkamp}, {Stegmann}, {Steppa}, {Takahashi}, {Tavernier},
  {Taylor}, {Terrier}, {Tiziani}, {Tluczykont}, {Trichard}, {Tsirou}, {Tsuji},
  {Tuffs}, {Uchiyama}, {van der Walt}, {van Eldik}, {van Rensburg}, {van
  Soelen}, {Vasileiadis}, {Veh}, {Venter}, {Vincent}, {Vink}, {Voisin},
  {V{\"o}lk}, {Vuillaume}, {Wadiasingh}, {Wagner}, {White}, {Wierzcholska},
  {Yang}, {Yoneda}, {Zacharias}, {Zanin}, {Zdziarski}, {Zech}, {Ziegler},
  {Zorn}, \& {{\.Z}ywucka}}]{2020Nature}
{H.~E.~S.~S. Collaboration}, {Abdalla}, H., {Adam}, R., {et~al.} 2020, \nat,
  582, 356, \dodoi{10.1038/s41586-020-2354-1}

\bibitem[{{Hahn}(2015)}]{hahn}
{Hahn}, J. 2015, in International Cosmic Ray Conference, Vol.~34, 34th
  International Cosmic Ray Conference (ICRC2015), 917

\bibitem[{{Heckman} {et~al.}(1983){Heckman}, {Lebofsky}, {Rieke}, \& {van
  Breugel}}]{heck}
{Heckman}, T.~M., {Lebofsky}, M.~J., {Rieke}, G.~H., \& {van Breugel}, W. 1983,
  \apj, 272, 400, \dodoi{10.1086/161308}

\bibitem[{{Ho}(2002)}]{ho02}
{Ho}, L.~C. 2002, \apj, 564, 120, \dodoi{10.1086/324399}

\bibitem[{{Ho}(2008)}]{ho08}
---. 2008, \araa, 46, 475, \dodoi{10.1146/annurev.astro.45.051806.110546}

\bibitem[{{Ho} {et~al.}(1995){Ho}, {Filippenko}, \& {Sargent}}]{ho95}
{Ho}, L.~C., {Filippenko}, A.~V., \& {Sargent}, W.~L. 1995, \apjs, 98, 477,
  \dodoi{10.1086/192170}

\bibitem[{{Ho} {et~al.}(1997){Ho}, {Filippenko}, {Sargent}, \& {Peng}}]{ho97}
{Ho}, L.~C., {Filippenko}, A.~V., {Sargent}, W. L.~W., \& {Peng}, C.~Y. 1997,
  \apjs, 112, 391, \dodoi{10.1086/313042}

\bibitem[{{Ho} {et~al.}(2000){Ho}, {Rudnick}, {Rix}, {Shields}, {McIntosh},
  {Filippenko}, {Sargent}, \& {Eracleous}}]{ho20}
{Ho}, L.~C., {Rudnick}, G., {Rix}, H.-W., {et~al.} 2000, \apj, 541, 120,
  \dodoi{10.1086/309440}

\bibitem[{{H{\"o}nig} \& {Beckert}(2007)}]{torus}
{H{\"o}nig}, S.~F., \& {Beckert}, T. 2007, \mnras, 380, 1172,
  \dodoi{10.1111/j.1365-2966.2007.12157.x}

\bibitem[{{Jaffe} {et~al.}(1993){Jaffe}, {Ford}, {Ferrarese}, {van den Bosch},
  \& {O'Connell}}]{jaffe}
{Jaffe}, W., {Ford}, H.~C., {Ferrarese}, L., {van den Bosch}, F., \&
  {O'Connell}, R.~W. 1993, \nat, 364, 213, \dodoi{10.1038/364213a0}

\bibitem[{{Jansen} {et~al.}(2001){Jansen}, {Lumb}, {Altieri}, {Clavel}, {Ehle},
  {Erd}, {Gabriel}, {Guainazzi}, {Gondoin}, {Much}, {Munoz}, {Santos},
  {Schartel}, {Texier}, \& {Vacanti}}]{jansen}
{Jansen}, F., {Lumb}, D., {Altieri}, B., {et~al.} 2001, \aap, 365, L1,
  \dodoi{10.1051/0004-6361:20000036}

\bibitem[{{Jones} \& {Wehrle}(1997)}]{jones}
{Jones}, D.~L., \& {Wehrle}, A.~E. 1997, \apj, 484, 186, \dodoi{10.1086/304320}

\bibitem[{{Kalberla} {et~al.}(2005){Kalberla}, {Burton}, {Hartmann}, {Arnal},
  {Bajaja}, {Morras}, \& {P{\"o}ppel}}]{karl}
{Kalberla}, P.~M.~W., {Burton}, W.~B., {Hartmann}, D., {et~al.} 2005, \aap,
  440, 775, \dodoi{10.1051/0004-6361:20041864}

\bibitem[{{Kormendy} \& {Ho}(2013)}]{korm13}
{Kormendy}, J., \& {Ho}, L.~C. 2013, \araa, 51, 511,
  \dodoi{10.1146/annurev-astro-082708-101811}

\bibitem[{{Kovalev} {et~al.}(2005){Kovalev}, {Kellermann}, {Lister}, {Homan},
  {Vermeulen}, {Cohen}, {Ros}, {Kadler}, {Lobanov}, {Zensus}, {Kardashev},
  {Gurvits}, {Aller}, \& {Aller}}]{kov}
{Kovalev}, Y.~Y., {Kellermann}, K.~I., {Lister}, M.~L., {et~al.} 2005, \aj,
  130, 2473, \dodoi{10.1086/497430}

\bibitem[{{Laor}(2003)}]{blr}
{Laor}, A. 2003, \apj, 590, 86, \dodoi{10.1086/375008}

\bibitem[{{Lazio} {et~al.}(2001){Lazio}, {Waltman}, {Ghigo}, {Fiedler},
  {Foster}, \& {Johnston}}]{lazio}
{Lazio}, T. J.~W., {Waltman}, E.~B., {Ghigo}, F.~D., {et~al.} 2001, \apjs, 136,
  265, \dodoi{10.1086/322531}

\bibitem[{{Lott} {et~al.}(2020){Lott}, {Gasparrini}, \& {Ciprini}}]{lott}
{Lott}, B., {Gasparrini}, D., \& {Ciprini}, S. 2020, arXiv e-prints,
  arXiv:2010.08406.
\newblock \doarXiv{2010.08406}

\bibitem[{{Lucchini} {et~al.}(2019){Lucchini}, {Krau{\ss}}, \&
  {Markoff}}]{lucchini}
{Lucchini}, M., {Krau{\ss}}, F., \& {Markoff}, S. 2019, \mnras, 489, 1633,
  \dodoi{10.1093/mnras/stz2125}

\bibitem[{{MAGIC Collaboration} {et~al.}(2020){MAGIC Collaboration}, {Acciari},
  {Ansoldi}, {Antonelli}, {Arbet Engels}, {Arcaro}, {Baack}, {Babi{\'c}},
  {Banerjee}, {Bangale}, {Barres de Almeida}, {Barrio}, {Becerra Gonz{\'a}lez},
  {Bednarek}, {Bellizzi}, {Bernardini}, {Berti}, {Besenrieder},
  {Bhattacharyya}, {Bigongiari}, {Biland}, {Blanch}, {Bonnoli},
  {Bo{\v{s}}njak}, {Busetto}, {Carosi}, {Ceribella}, {Chai}, {Chilingaryan},
  {Cikota}, {Colak}, {Colin}, {Colombo}, {Contreras}, {Cortina}, {Covino},
  {D'Elia}, {da Vela}, {Dazzi}, {de Angelis}, {de Lotto}, {Delfino}, {Delgado},
  {Depaoli}, {di Pierro}, {di Venere}, {Do Souto Espi{\~n}eira}, {Dominis
  Prester}, {Donini}, {Dorner}, {Doro}, {Elsaesser}, {Fallah Ramazani},
  {Fattorini}, {Fern{\'a}ndez-Barral}, {Ferrara}, {Fidalgo}, {Foffano},
  {Fonseca}, {Font}, {Fruck}, {Fukami}, {Garc{\'\i}a L{\'o}pez}, {Garczarczyk},
  {Gasparyan}, {Gaug}, {Giglietto}, {Giordano}, {Godinovi{\'c}}, {Green},
  {Guberman}, {Hadasch}, {Hahn}, {Herrera}, {Hoang}, {Hrupec}, {H{\"u}tten},
  {Inada}, {Inoue}, {Ishio}, {Iwamura}, {Jouvin}, {Kerszberg}, {Kubo},
  {Kushida}, {Lamastra}, {Lelas}, {Leone}, {Lindfors}, {Lombardi}, {Longo},
  {L{\'o}pez}, {L{\'o}pez-Coto}, {L{\'o}pez-Oramas}, {Loporchio}, {Machado de
  Oliveira Fraga}, {Maggio}, {Majumdar}, {Makariev}, {Mallamaci}, {Maneva},
  {Manganaro}, {Mannheim}, {Maraschi}, {Mariotti}, {Mart{\'\i}nez}, {Masuda},
  {Mazin}, {Mi{\'c}anovi{\'c}}, {Miceli}, {Minev}, {Miranda}, {Mirzoyan},
  {Molina}, {Moralejo}, {Morcuende}, {Moreno}, {Moretti}, {Munar-Adrover},
  {Neustroev}, {Nigro}, {Nilsson}, {Ninci}, {Nishijima}, {Noda}, {Nogu{\'e}s},
  {N{\"o}the}, {Nozaki}, {Paiano}, {Palacio}, {Palatiello}, {Paneque},
  {Paoletti}, {Paredes}, {Pe{\~n}il}, {Peresano}, {Persic}, {Prada Moroni},
  {Prandini}, {Puljak}, {Rhode}, {Rib{\'o}}, {Rico}, {Righi}, {Rugliancich},
  {Saha}, {Sahakyan}, {Saito}, {Sakurai}, {Satalecka}, {Schmidt}, {Schweizer},
  {Sitarek}, {{\v{S}}nidari{\'c}}, {Sobczynska}, {Somero}, {Stamerra}, {Strom},
  {Strzys}, {Suda}, {Suri{\'c}}, {Takahashi}, {Tavecchio}, {Temnikov},
  {Terzi{\'c}}, {Teshima}, {Torres-Alb{\`a}}, {Tosti}, {Tsujimoto}, {Vagelli},
  {van Scherpenberg}, {Vanzo}, {Acosta}, {Vigorito}, {Vitale}, {Vovk}, {Will},
  {Zari{\'c}}, {Asano}, {Hada}, {Harris}, {Giroletti}, {Jermak}, {Madrid},
  {Massaro}, {Richter}, {Spanier}, {Steele}, \& {Walker}}]{magic}
{MAGIC Collaboration}, {Acciari}, V.~A., {Ansoldi}, S., {et~al.} 2020, \mnras,
  492, 5354, \dodoi{10.1093/mnras/staa014}

\bibitem[{{Magorrian} {et~al.}(1998){Magorrian}, {Tremaine}, {Richstone},
  {Bender}, {Bower}, {Dressler}, {Faber}, {Gebhardt}, {Green}, {Grillmair},
  {Kormendy}, \& {Lauer}}]{magorrian}
{Magorrian}, J., {Tremaine}, S., {Richstone}, D., {et~al.} 1998, \aj, 115,
  2285, \dodoi{10.1086/300353}

\bibitem[{{Maraschi} {et~al.}(1992){Maraschi}, {Ghisellini}, \&
  {Celotti}}]{maras}
{Maraschi}, L., {Ghisellini}, G., \& {Celotti}, A. 1992, \apjl, 397, L5,
  \dodoi{10.1086/186531}

\bibitem[{{Marinelli} {et~al.}(2014){Marinelli}, {Fraija}, \&
  {Patricelli}}]{2014marinelli}
{Marinelli}, A., {Fraija}, N., \& {Patricelli}, B. 2014, arXiv e-prints,
  arXiv:1410.8549.
\newblock \doarXiv{1410.8549}

\bibitem[{{Marshall} {et~al.}(2002){Marshall}, {Miller}, {Davis}, {Perlman},
  {Wise}, {Canizares}, \& {Harris}}]{marshall02}
{Marshall}, H.~L., {Miller}, B.~P., {Davis}, D.~S., {et~al.} 2002, \apj, 564,
  683, \dodoi{10.1086/324396}

\bibitem[{{Massaro} {et~al.}(2004){Massaro}, {Perri}, {Giommi}, \&
  {Nesci}}]{massaro}
{Massaro}, E., {Perri}, M., {Giommi}, P., \& {Nesci}, R. 2004, \aap, 413, 489,
  \dodoi{10.1051/0004-6361:20031558}

\bibitem[{{Mattox} {et~al.}(1996){Mattox}, {Bertsch}, {Chiang}, {Dingus},
  {Digel}, {Esposito}, {Fierro}, {Hartman}, {Hunter}, {Kanbach}, {Kniffen},
  {Lin}, {Macomb}, {Mayer-Hasselwander}, {Michelson}, {von Montigny},
  {Mukherjee}, {Nolan}, {Ramanamurthy}, {Schneid}, {Sreekumar}, {Thompson}, \&
  {Willis}}]{likeli}
{Mattox}, J.~R., {Bertsch}, D.~L., {Chiang}, J., {et~al.} 1996, \apj, 461, 396,
  \dodoi{10.1086/177068}

\bibitem[{{Merloni} {et~al.}(2003){Merloni}, {Heinz}, \& {di Matteo}}]{merloni}
{Merloni}, A., {Heinz}, S., \& {di Matteo}, T. 2003, \mnras, 345, 1057,
  \dodoi{10.1046/j.1365-2966.2003.07017.x}

\bibitem[{{Mezcua} \& {Prieto}(2014)}]{mez}
{Mezcua}, M., \& {Prieto}, M.~A. 2014, \apj, 787, 62,
  \dodoi{10.1088/0004-637X/787/1/62}

\bibitem[{{Mukherjee} \& {VERITAS Collaboration}(2017)}]{mukh}
{Mukherjee}, R., \& {VERITAS Collaboration}. 2017, The Astronomer's Telegram,
  9931, 1

\bibitem[{{Nagai} {et~al.}(2010){Nagai}, {Suzuki}, {Asada}, {Kino}, {Kameno},
  {Doi}, {Inoue}, {Kataoka}, {Bach}, {Hirota}, {Matsumoto}, {Honma},
  {Kobayashi}, \& {Fujisawa}}]{nagai10}
{Nagai}, H., {Suzuki}, K., {Asada}, K., {et~al.} 2010, \pasj, 62, L11,
  \dodoi{10.1093/pasj/62.2.L11}

\bibitem[{{Nagar} {et~al.}(2005){Nagar}, {Falcke}, \& {Wilson}}]{nagar05}
{Nagar}, N.~M., {Falcke}, H., \& {Wilson}, A.~S. 2005, \aap, 435, 521,
  \dodoi{10.1051/0004-6361:20042277}

\bibitem[{{Nagar} {et~al.}(2001){Nagar}, {Wilson}, \& {Falcke}}]{nag1}
{Nagar}, N.~M., {Wilson}, A.~S., \& {Falcke}, H. 2001, \apjl, 559, L87,
  \dodoi{10.1086/323938}

\bibitem[{{Narayan} \& {McClintock}(2008)}]{nar_mc}
{Narayan}, R., \& {McClintock}, J.~E. 2008, \nar, 51, 733,
  \dodoi{10.1016/j.newar.2008.03.002}

\bibitem[{{Narayan} \& {Yi}(1994)}]{narayan}
{Narayan}, R., \& {Yi}, I. 1994, \apjl, 428, L13, \dodoi{10.1086/187381}

\bibitem[{{Nemmen} {et~al.}(2007){Nemmen}, {Bower}, {Babul}, \&
  {Storchi-Bergmann}}]{nemmen07}
{Nemmen}, R.~S., {Bower}, R.~G., {Babul}, A., \& {Storchi-Bergmann}, T. 2007,
  \mnras, 377, 1652, \dodoi{10.1111/j.1365-2966.2007.11726.x}

\bibitem[{{Nemmen} {et~al.}(2014){Nemmen}, {Storchi-Bergmann}, \&
  {Eracleous}}]{nemmen}
{Nemmen}, R.~S., {Storchi-Bergmann}, T., \& {Eracleous}, M. 2014, \mnras, 438,
  2804, \dodoi{10.1093/mnras/stt2388}

\bibitem[{{Nemmen} {et~al.}(2010){Nemmen}, {Storchi-Bergmann}, {Eracleous}, \&
  {Yuan}}]{nemmen10}
{Nemmen}, R.~S., {Storchi-Bergmann}, T., {Eracleous}, M., \& {Yuan}, F. 2010,
  in Co-Evolution of Central Black Holes and Galaxies, ed. B.~M. {Peterson},
  R.~S. {Somerville}, \& T.~{Storchi-Bergmann}, Vol. 267, 313--318,
  \dodoi{10.1017/S1743921310006538}

\bibitem[{{Perlman} {et~al.}(2001){Perlman}, {Sparks}, {Radomski}, {Packham},
  {Fisher}, {Pi{\~n}a}, \& {Biretta}}]{perl}
{Perlman}, E.~S., {Sparks}, W.~B., {Radomski}, J., {et~al.} 2001, \apjl, 561,
  L51, \dodoi{10.1086/324515}

\bibitem[{{Piner} {et~al.}(2001){Piner}, {Jones}, \& {Wehrle}}]{piner}
{Piner}, B.~G., {Jones}, D.~L., \& {Wehrle}, A.~E. 2001, \aj, 122, 2954,
  \dodoi{10.1086/323927}

\bibitem[{{Prieto} {et~al.}(2016){Prieto}, {Fern{\'a}ndez-Ontiveros},
  {Markoff}, {Espada}, \& {Gonz{\'a}lez-Mart{\'\i}n}}]{prieto}
{Prieto}, M.~A., {Fern{\'a}ndez-Ontiveros}, J.~A., {Markoff}, S., {Espada}, D.,
  \& {Gonz{\'a}lez-Mart{\'\i}n}, O. 2016, \mnras, 457, 3801,
  \dodoi{10.1093/mnras/stw166}

\bibitem[{{Quataert} {et~al.}(1999){Quataert}, {Di Matteo}, {Narayan}, \&
  {Ho}}]{quat}
{Quataert}, E., {Di Matteo}, T., {Narayan}, R., \& {Ho}, L.~C. 1999, \apjl,
  525, L89, \dodoi{10.1086/312353}

\bibitem[{{Reynolds} {et~al.}(1996){Reynolds}, {Di Matteo}, {Fabian}, {Hwang},
  \& {Canizares}}]{reynolds}
{Reynolds}, C.~S., {Di Matteo}, T., {Fabian}, A.~C., {Hwang}, U., \&
  {Canizares}, C.~R. 1996, \mnras, 283, L111, \dodoi{10.1093/mnras/283.4.L111}

\bibitem[{{Rodrigues} {et~al.}(2021){Rodrigues}, {Heinze}, {Palladino}, {van
  Vliet}, \& {Winter}}]{rodrigues}
{Rodrigues}, X., {Heinze}, J., {Palladino}, A., {van Vliet}, A., \& {Winter},
  W. 2021, \prl, 126, 191101, \dodoi{10.1103/PhysRevLett.126.191101}

\bibitem[{{Sikora} {et~al.}(2007){Sikora}, {Stawarz}, \& {Lasota}}]{sikora}
{Sikora}, M., {Stawarz}, {\L}., \& {Lasota}, J.-P. 2007, \apj, 658, 815,
  \dodoi{10.1086/511972}

\bibitem[{{Stawarz} {et~al.}(2006){Stawarz}, {Kneiske}, \&
  {Kataoka}}]{stawarz06}
{Stawarz}, {\L}., {Kneiske}, T.~M., \& {Kataoka}, J. 2006, \apj, 637, 693,
  \dodoi{10.1086/498084}

\bibitem[{{Stawarz} {et~al.}(2003){Stawarz}, {Sikora}, \&
  {Ostrowski}}]{stawarz03}
{Stawarz}, {\L}., {Sikora}, M., \& {Ostrowski}, M. 2003, \apj, 597, 186,
  \dodoi{10.1086/378290}

\bibitem[{{Str{\"u}der} {et~al.}(2001){Str{\"u}der}, {Briel}, {Dennerl},
  {Hartmann}, {Kendziorra}, {Meidinger}, {Pfeffermann}, {Reppin}, {Aschenbach},
  {Bornemann}, {Br{\"a}uninger}, {Burkert}, {Elender}, {Freyberg}, {Haberl},
  {Hartner}, {Heuschmann}, {Hippmann}, {Kastelic}, {Kemmer}, {Kettenring},
  {Kink}, {Krause}, {M{\"u}ller}, {Oppitz}, {Pietsch}, {Popp}, {Predehl},
  {Read}, {Stephan}, {St{\"o}tter}, {Tr{\"u}mper}, {Holl}, {Kemmer}, {Soltau},
  {St{\"o}tter}, {Weber}, {Weichert}, {von Zanthier}, {Carathanassis}, {Lutz},
  {Richter}, {Solc}, {B{\"o}ttcher}, {Kuster}, {Staubert}, {Abbey}, {Holland},
  {Turner}, {Balasini}, {Bignami}, {La Palombara}, {Villa}, {Buttler},
  {Gianini}, {Lain{\'e}}, {Lumb}, \& {Dhez}}]{struder}
{Str{\"u}der}, L., {Briel}, U., {Dennerl}, K., {et~al.} 2001, \aap, 365, L18,
  \dodoi{10.1051/0004-6361:20000066}

\bibitem[{{Sun} {et~al.}(2018){Sun}, {Yang}, {Rieger}, {Liu}, \&
  {Aharonian}}]{2018sun}
{Sun}, X.-N., {Yang}, R.-Z., {Rieger}, F.~M., {Liu}, R.-Y., \& {Aharonian}, F.
  2018, \aap, 612, A106, \dodoi{10.1051/0004-6361/201731716}

\bibitem[{{Tanada} {et~al.}(2018{\natexlab{a}}){Tanada}, {Kataoka}, {Arimoto},
  {Akita}, {Cheung}, {Digel}, \& {Fukazawa}}]{tanada18}
{Tanada}, K., {Kataoka}, J., {Arimoto}, M., {et~al.} 2018{\natexlab{a}}, \apj,
  860, 74, \dodoi{10.3847/1538-4357/aac26b}

\bibitem[{{Tanada} {et~al.}(2018{\natexlab{b}}){Tanada}, {Kataoka}, {Arimoto},
  {Akita}, {Cheung}, {Digel}, \& {Fukazawa}}]{tanada_chandra}
---. 2018{\natexlab{b}}, \apj, 860, 74, \dodoi{10.3847/1538-4357/aac26b}

\bibitem[{Terashima {et~al.}(2000)Terashima, Ho, \& Ptak}]{Terashima_2000}
Terashima, Y., Ho, L.~C., \& Ptak, A.~F. 2000, The Astrophysical Journal, 539,
  161, \dodoi{10.1086/309234}

\bibitem[{{The Event Horizon Collaboration}(2021)}]{eht21}
{The Event Horizon Collaboration}. 2021, arXiv e-prints, arXiv:2105.01173.
\newblock \doarXiv{2105.01173}

\bibitem[{{Trager} {et~al.}(2000){Trager}, {Faber}, {Worthey}, \&
  {Gonz{\'a}lez}}]{trager}
{Trager}, S.~C., {Faber}, S.~M., {Worthey}, G., \& {Gonz{\'a}lez}, J.~J. 2000,
  \aj, 119, 1645, \dodoi{10.1086/301299}

\bibitem[{{Turner} {et~al.}(2001){Turner}, {Abbey}, {Arnaud}, {Balasini},
  {Barbera}, {Belsole}, {Bennie}, {Bernard}, {Bignami}, {Boer}, {Briel},
  {Butler}, {Cara}, {Chabaud}, {Cole}, {Collura}, {Conte}, {Cros}, {Denby},
  {Dhez}, {Di Coco}, {Dowson}, {Ferrando}, {Ghizzardi}, {Gianotti}, {Goodall},
  {Gretton}, {Griffiths}, {Hainaut}, {Hochedez}, {Holland}, {Jourdain},
  {Kendziorra}, {Lagostina}, {Laine}, {La Palombara}, {Lortholary}, {Lumb},
  {Marty}, {Molendi}, {Pigot}, {Poindron}, {Pounds}, {Reeves}, {Reppin},
  {Rothenflug}, {Salvetat}, {Sauvageot}, {Schmitt}, {Sembay}, {Short},
  {Spragg}, {Stephen}, {Str{\"u}der}, {Tiengo}, {Trifoglio}, {Tr{\"u}mper},
  {Vercellone}, {Vigroux}, {Villa}, {Ward}, {Whitehead}, \& {Zonca}}]{turner}
{Turner}, M.~J.~L., {Abbey}, A., {Arnaud}, M., {et~al.} 2001, \aap, 365, L27,
  \dodoi{10.1051/0004-6361:20000087}

\bibitem[{{Venturi} {et~al.}(1993){Venturi}, {Giovannini}, {Feretti},
  {Comoretto}, \& {Wehrle}}]{venturi}
{Venturi}, T., {Giovannini}, G., {Feretti}, L., {Comoretto}, G., \& {Wehrle},
  A.~E. 1993, \apj, 408, 81, \dodoi{10.1086/172571}

\bibitem[{{Verdoes Kleijn} {et~al.}(1999){Verdoes Kleijn}, {Baum}, {de Zeeuw},
  \& {O'Dea}}]{verdoes99}
{Verdoes Kleijn}, G.~A., {Baum}, S.~A., {de Zeeuw}, P.~T., \& {O'Dea}, C.~P.
  1999, \aj, 118, 2592, \dodoi{10.1086/301135}

\bibitem[{{Verdoes Kleijn} {et~al.}(2002){Verdoes Kleijn}, {Baum}, {de Zeeuw},
  \& {O'Dea}}]{verdoes}
---. 2002, \aj, 123, 1334, \dodoi{10.1086/339177}

\bibitem[{{Vermeulen} {et~al.}(1994){Vermeulen}, {Readhead}, \&
  {Backer}}]{verm}
{Vermeulen}, R.~C., {Readhead}, A.~C.~S., \& {Backer}, D.~C. 1994, \apjl, 430,
  L41, \dodoi{10.1086/187433}

\bibitem[{{Walker} {et~al.}(2000){Walker}, {Dhawan}, {Romney}, {Kellermann}, \&
  {Vermeulen}}]{walker00}
{Walker}, R.~C., {Dhawan}, V., {Romney}, J.~D., {Kellermann}, K.~I., \&
  {Vermeulen}, R.~C. 2000, \apj, 530, 233, \dodoi{10.1086/308372}

\bibitem[{{Walker} {et~al.}(2018){Walker}, {Hardee}, {Davies}, {Ly}, \&
  {Junor}}]{walker}
{Walker}, R.~C., {Hardee}, P.~E., {Davies}, F.~B., {Ly}, C., \& {Junor}, W.
  2018, \apj, 855, 128, \dodoi{10.3847/1538-4357/aaafcc}

\bibitem[{{Walker} {et~al.}(1994){Walker}, {Romney}, \& {Benson}}]{walker94}
{Walker}, R.~C., {Romney}, J.~D., \& {Benson}, J.~M. 1994, \apjl, 430, L45,
  \dodoi{10.1086/187434}

\bibitem[{{Whysong} \& {Antonucci}(2004)}]{why}
{Whysong}, D., \& {Antonucci}, R. 2004, \apj, 602, 116, \dodoi{10.1086/380828}

\bibitem[{{Wilms} {et~al.}(2000){Wilms}, {Allen}, \& {McCray}}]{wilms00}
{Wilms}, J., {Allen}, A., \& {McCray}, R. 2000, \apj, 542, 914,
  \dodoi{10.1086/317016}

\bibitem[{{Wood} {et~al.}(2017{\natexlab{a}}){Wood}, {Caputo}, {Charles}, {Di
  Mauro}, {Magill}, {Perkins}, \& {Fermi-LAT Collaboration}}]{fermipy}
{Wood}, M., {Caputo}, R., {Charles}, E., {et~al.} 2017{\natexlab{a}}, in
  International Cosmic Ray Conference, Vol. 301, 35th International Cosmic Ray
  Conference (ICRC2017), 824.
\newblock \doarXiv{1707.09551}

\bibitem[{{Wood} {et~al.}(2017{\natexlab{b}}){Wood}, {Caputo}, {Charles}, {Di
  Mauro}, {Magill}, {Perkins}, \& {Fermi-LAT Collaboration}}]{wood_fermipy}
{Wood}, M., {Caputo}, R., {Charles}, E., {et~al.} 2017{\natexlab{b}}, in
  International Cosmic Ray Conference, Vol. 301, 35th International Cosmic Ray
  Conference (ICRC2017), 824.
\newblock \doarXiv{1707.09551}

\bibitem[{{Worrall} {et~al.}(2003){Worrall}, {Birkinshaw}, \&
  {Hardcastle}}]{wor}
{Worrall}, D.~M., {Birkinshaw}, M., \& {Hardcastle}, M.~J. 2003, \mnras, 343,
  L73, \dodoi{10.1046/j.1365-8711.2003.06945.x}

\bibitem[{{Worrall} {et~al.}(2007{\natexlab{a}}){Worrall}, {Birkinshaw},
  {Laing}, {Cotton}, \& {Bridle}}]{worrall315}
{Worrall}, D.~M., {Birkinshaw}, M., {Laing}, R.~A., {Cotton}, W.~D., \&
  {Bridle}, A.~H. 2007{\natexlab{a}}, \mnras, 380, 2,
  \dodoi{10.1111/j.1365-2966.2007.11998.x}

\bibitem[{{Worrall} {et~al.}(2007{\natexlab{b}}){Worrall}, {Birkinshaw},
  {Laing}, {Cotton}, \& {Bridle}}]{worall}
---. 2007{\natexlab{b}}, \mnras, 380, 2,
  \dodoi{10.1111/j.1365-2966.2007.11998.x}

\bibitem[{{Worrall} {et~al.}(2010){Worrall}, {Birkinshaw}, {O'Sullivan},
  {Zezas}, {Wolter}, {Trinchieri}, \& {Fabbiano}}]{worrall4261}
{Worrall}, D.~M., {Birkinshaw}, M., {O'Sullivan}, E., {et~al.} 2010, \mnras,
  408, 701, \dodoi{10.1111/j.1365-2966.2010.17162.x}

\bibitem[{{Younes} {et~al.}(2012){Younes}, {Porquet}, {Sabra}, {Reeves}, \&
  {Grosso}}]{younes}
{Younes}, G., {Porquet}, D., {Sabra}, B., {Reeves}, J.~N., \& {Grosso}, N.
  2012, \aap, 539, A104, \dodoi{10.1051/0004-6361/201118299}

\bibitem[{{Young} {et~al.}(1995){Young}, {Xie}, {Tacconi}, {Knezek}, {Viscuso},
  {Tacconi-Garman}, {Scoville}, {Schneider}, {Schloerb}, {Lord}, {Lesser},
  {Kenney}, {Huang}, {Devereux}, {Claussen}, {Case}, {Carpenter}, {Berry}, \&
  {Allen}}]{young}
{Young}, J.~S., {Xie}, S., {Tacconi}, L., {et~al.} 1995, \apjs, 98, 219,
  \dodoi{10.1086/192159}

\bibitem[{{Yu} {et~al.}(2011){Yu}, {Yuan}, \& {Ho}}]{yu11}
{Yu}, Z., {Yuan}, F., \& {Ho}, L.~C. 2011, \apj, 726, 87,
  \dodoi{10.1088/0004-637X/726/2/87}

\bibitem[{{Zacharias} \& {Wagner}(2016)}]{zacharias}
{Zacharias}, M., \& {Wagner}, S. 2016, Galaxies, 4, 63,
  \dodoi{10.3390/galaxies4040063}

\bibitem[{{Zezas} {et~al.}(2005){Zezas}, {Birkinshaw}, {Worrall}, {Peters}, \&
  {Fabbiano}}]{zez05}
{Zezas}, A., {Birkinshaw}, M., {Worrall}, D.~M., {Peters}, A., \& {Fabbiano},
  G. 2005, \apj, 627, 711, \dodoi{10.1086/430044}

\end{thebibliography}
\bibliographystyle{aasjournal}



\end{document}